\documentclass[prb,twocolumn,showpacs,showkeys,superscriptaddress]{revtex4-1}
\usepackage{amsmath}    
\usepackage{amsfonts}
\usepackage{amssymb}
\usepackage{graphicx}   
\usepackage{epstopdf}
\usepackage{float} 
\usepackage[caption=false]{subfig} 
\usepackage{color} 
\usepackage[normalem]{ulem} 
\usepackage[makeroom]{cancel} 
\usepackage{hyperref} 
\usepackage{siunitx} 
\hypersetup{colorlinks, citecolor=black, filecolor=black, linkcolor=black, urlcolor=black}

\DeclareMathOperator{\erf}{erf}
\DeclareMathOperator{\const}{const}

\begin{document}

\title{Polarizability of electrically induced magnetic vortex plasma}
\author{P.I.\ Karpov}
  \email{karpov.petr@gmail.com}   
\author{S.I.\ Mukhin}

\affiliation{Department of Theoretical Physics and Quantum Technologies, National University of Science and Technology MISiS,  Leninski avenue 4, 119991, Moscow, Russia}
\date{\today}

\begin{abstract}
Electric field control of magnetic structures, particularly topological defects in magnetoelectric materials, draws a great attention in recent years, which has led to experimental success in creation and manipulation by electric field of single magnetic defects, such as domain walls and skyrmions.
In this work we explore a scenario of electric field creation of another type of topological defects -- magnetic vortices and antivortices, which are characteristic for materials with easy plane (XY) symmetry.
Each magnetic (anti)vortex in magnetoelectric materials (such as type-II multiferroics) possesses a quantized magnetic and an electric charge, where the former is responsible for interaction between vortices and the latter couples the vortices to electric field. This property of magnetic vortices opens a peculiar possibility of creation of magnetic vortex  plasma by non-uniform electric fields.
We show that the electric field, created by a cantilever tip, produces a ``magnetic atom'' with a localized spatially ordered spot of vortices (``nucleus'' of the atom) surrounded by antivortices (``electronic shells'').
We analytically find the vortex density distribution profile and temperature dependence of polarizability of this structure and confirm it numerically.
We show that electric polarizability of the ``magnetic atom'' depends on temperature as $\alpha \sim 1/T^{1-\eta}$ ($\eta>0$), which is consistent with Euclidean random matrix theory prediction.
\end{abstract}

\maketitle


\section{Introduction}

Efficient control of magnetic structures by electric fields might revolutionize the field of magnetic memory devices.
Compared to the existing technology that switches magnetic states using electric currents (via created magnetic fields or spin-transfer torques) and thus is subject to joule heating \cite{Taniyama:2015}, the replacement of current with an electric field promises to reduce energy dissipation by the factor of 100 [\onlinecite{Matsukura:2015}]. Several mechanisms of the electric-field manipulation of magnetism were proposed such as: electric-field changing the carriers concentration, which mediate magnetic interactions in magnetic semiconductors; changing the magnetic anisotropy in insulator/magnetic metal heterostructures; using magnetoelectric coupling in multiferroic materials \cite{Matsukura:2015}.
In the present work we will focus on the latter mechanism.

Magnetoelectric effect was first discovered ``with the point of the pen'' by Dzyaloshinskii \cite{Dzyaloshinskii:1959} and then this prediction was experimentally observed by Astrov in $\mathrm{Cr_2 O_3}$ [\onlinecite{Astrov:1960}]. Then, after a fast development in 60s-70s, interest to such materials gradually faded away, because of the weakness of the observed effect. But since the experimental successes of 2000-s  -- discovery of many multiferroics (materials in which two or more ferroic orders, such as  ferromagnetism and ferroelectricity, are present \cite{Schmid:1994}) and ``revival'' of the field, magnetoelectric materials are still in the focus of both experimental and theoretical research. For recent reviews on multiferroics see \cite{Khomskii:2016,Fiebig:2016,Tokura:2014,Zvezdin:2012,Spaldin:2010}.

According to the classification of Khomskii \cite{Khomskii:2009}, there are two types of multiferroics. Type I consists of multiferroics, where both ferromagnetic and ferroelectric orders coexist, being almost not coupled to each other. While in type-II multiferroics the two orders are coupled and usually a magnetic order creates an electric one. This feature can be used backwards: applying an electric field and inducing polarization, magnetic patterns can be also created and manipulated. While this type-II feature was originally coined to multiferroics, it becomes more spread that this coupling between magnetization and electric polarization can appear in any magnetic insulator, where it is allowed by symmetry \cite{Khomskii:2016}. In 2008 Dzyaloshinskii theoretically predicted the possibility of creation and manipulation of domain walls by electric fields in conventional and weak ferromagnets  \cite{Dzyaloshinskii:2008} and the movement of magnetic domain walls under the action of electric field was directly observed \cite{Logginov:2008,Pyatakov:2011}.

The use of domain walls and other topological defects in multiferroics \cite{Seidel:2016} for magnetic storage is particularly appealing since it provides a mechanism of very dense packing of information, while topological nature protects information from loss under the influence of external perturbations such as heating or mechanical action \cite{Parkin:2008,Fert:2013}.
Among the bright examples of such topological defects are skyrmion lattices \cite{Tokura:2012}, as well as individual skyrmions observed in magnetoelectric materials \cite{Hsu:2016}, magnetic vortices \cite{Mostovoy:2009,Pyatakov:2012} and their discrete analogues \cite{Cheong:2010,Mostovoy:2010,Artyukhin:2014}, domain walls \cite{Logginov:2008,Pyatakov:2011, Meier:2015}. Usually such defects and textures are observed in thin-film magnetoelectric materials \cite{Ramesh:2007}.

In the present paper we consider theoretically a phenomenological model developed by Mostovoy \cite{Mostovoy:2006} for a type-II multiferroics and apply it to a thin-film material with easy plane symmetry. We show both analytically and numerically that in this case a strong enough locally-concentrated electric field  may cause formation of a ``magnetic atom'' that consists of vortex ``nucleus'' surrounded by antivortex ``shells''. This idea allows us to propose an experimental realization for 2D Coulomb plasma in a trap and make predictions about its normal modes. The topic of finite Coulomb  plasma clusters, and particularly of their normal modes, draws great attention in different areas of physics (see [\onlinecite{Melzer:2001}, \onlinecite{Melzer:2012}] and references therein).
However, usually particles in 2D plasma clusters interact between themselves via screened or unscreened 3D Coulomb $1/r$ potential. On the contrary, systems that we consider in this work provide a unique possibility of creation of plasma with constituents interacting via 2D Coulomb $\log r$ potential.

Also presented here research has some reminiscence with quantum electrodynamics. It was proposed that space- or time- dependent electric fields can generate electron-positron pairs from vacuum or fill empty electronic shells in an atom, producing a positron \cite{Kleinert:2008}. The idea was first theoretically proposed by Sauter [\onlinecite{Sauter:1931}] and Schwinger [\onlinecite{Schwinger:1951}] and attracts considerable interest in the present days (see, for example, [\onlinecite{Kleinert:2013}]). But the experimental observation is still lacking: such strong electric fields can not be obtained in laboratory conditions yet; now steps are being done in this direction: Extreme Light Infrastructure project \cite{ELI}, which will presumably allow to reach Schwinger limit \cite{Dunne:2009}, is currently under construction.

Another example lies in the physics of heavy atoms. For atomic numbers $Z \gtrsim 1/\alpha$ ($\alpha \approx 1/137$ is the fine structure constant) electric field becomes so strong that the nucleus can tear electrons out of vacuum and fill lower electron levels while emitting positrons \cite{Pomeranchuk:1945, Zeldovich:1971}. This effect was not yet observed since such heavy atoms are very unstable. However, recently a condensed-matter version of the artificial atom-like structures was realized in graphene \cite{Mao:2016}.

The present paper proposes an experimental setup for observation of analogous effects in another part of the realm of condensed matter systems, in multiferroics: we propose that vortex-antivortex pairs can be created from the ground state of a type-II multiferroic thin film by applying a strong non-uniform electric field.

The paper is organized as follows. In Sec.II we introduce our basic phenomenological model of quasi-two-dimensional type-II multiferroic-like material. In Sec.III we explore the consequences of applying strong non-uniform electric field created by a tip of cantilever and find ``the first critical electric field'', when the first vortex-antivortex pair is created. In Sec.IV we present our analytical results on the number of induced vortices and vortex density distribution in the continuous model. In Sec.V we deal explicitly with the effects of discreteness of vortex distribution  and estimate the principal part of temperature dependence of polarizability of artificial ``magnetic atom''.  Sec.VI presents our numerical results, which confirm theoretical calculations in Sec. IV and V.
In Sec.VII we make some estimates for the real materials and show that the critical electric field may be quite moderate and that even weak magnetic anisotropy protects in-plane spin arrangement.
Sec.VIII contains our conclusions and some ideas on experimental techniques that can be applied for observation of the described effects.


\section{Derivation of the model}
\label{Sec_Model}

In order to describe a 2D material with coupled magnetic and electric subsystems we use the following phenomenological model.
We write the total free energy density as a sum of parts arising from electric polarization, magnetization, and magnetoelectric coupling respectively:
\begin{align}
w = w_e + w_m + w_{me}.
\end{align}
Here
\begin{align}
w_e = \frac{P^2}{2\chi_e} - {\mathbf E} {\mathbf P} - \frac{{\mathbf E}^2}{8\pi}
\label{w_e}
\end{align}
is the energy of electric polarization \cite{LL8}, $\chi_e$ is the electric susceptibility in the absence of magnetization; we assume that there is no spontaneous polarization in the absence of magnetization, thus the $\sim P^{4}$ term is absent in the free energy (\ref{w_e}).
We write the magnetic part of the free energy as
\begin{align}
\label{w_m}
w_m = \frac{\alpha}{2} \left[ \left( \frac{\partial {\mathbf M}}{\partial x} \right)^2 +
\left(\frac{\partial {\mathbf M}}{\partial y} \right)^2 \right]
\end{align}
which is magnetic non-uniformity energy \cite{LL8}; contribution $A M^2 + B M^4$ that keeps $|{\mathbf M}| = M_0 = const$ is implied.
Since we are modeling thin-film material, we assume that the film thickness $h$ is much less than the exchange length and neglect the variations of ${\mathbf M}$ in the transverse to the film direction.
We also note that the same kind of analysis can be done also for an antiferromagnetic model, which also possesses the vortex excitations \cite{Yanagisawa:2013}.

The most interesting energy part is the coupling energy \cite{Mostovoy:2006}, which characterizes type-II multiferroics
\begin{align}
\label{w_me}
w_{me}= \gamma{\mathbf P}({\mathbf M}(\nabla{\mathbf M})-({\mathbf M}\nabla){\mathbf M}).
\end{align}
This Lifshits-type term was introduced based on symmetry grounds; it can be present if spatial inversion symmetry is broken. One of the possible underlying microscopic mechanisms is based on the inverse Dzyaloshinskii-Moriya interaction \cite{Katsura:2005}.

Combining three contributions (\ref{w_e}), (\ref{w_m}), (\ref{w_me}) we write the total free energy density of the magnetoelectric material as
\begin{align}
\label{energy-density}
w =  \frac{P^2}{2 \chi_e} - {\mathbf E}{\mathbf P} - \frac{{\mathbf E}^2}{8\pi} - \gamma{\mathbf P}(({\mathbf M}\nabla){\mathbf M} - {\mathbf M}(\nabla{\mathbf M})) + \nonumber\\
+\frac{\alpha}{2} \left[ \left( \frac{\partial {\mathbf M}}{\partial x} \right)^2 +
\left( \frac{\partial {\mathbf M}}{\partial y} \right)^2 \right].
\end{align}
Now we assume that ${\mathbf M}$ is a primary order parameter, which can induce ${\mathbf P}$. So that we can find polarization by minimizing  (\ref{energy-density}) with respect to $\mathbf P$
\begin{equation}
\label{PM}
\mathbf P = \gamma\chi_e (({\mathbf M}\nabla){\mathbf M} - {\mathbf M}(\nabla{\mathbf M})) + \chi_e {\mathbf E}.
\end{equation}
Because of the thin film geometry of the sample, magnetization tends to be parallel to the film plane, otherwise the magnetic energy of the demagnetizing field would greatly increase; also material can possess additional easy-plane type of symmetry. Therefore we restrict the magnetization to lie in the plane
\begin{equation}
\label{M}
{\mathbf M} (\mathbf{r}) = M_0
\left(
\begin{array}{ccc}
 \cos \phi(\mathbf{r}) \\
 \sin \phi(\mathbf{r}) \\
 0 \\
 \end{array}
\right).
\end{equation}
%
Using the angle variable $\phi$ the polarization can be written as
\begin{equation}
\label{polarization}
{\mathbf P} = \gamma\chi_e M_0^2  \left(
\begin{array}{ccc}
- \partial_y\phi \\
  \partial_x\phi \\
  0 \\
 \end{array}
\right) + \chi_e {\mathbf E}.
\end{equation}
Substituting (\ref{M}) and (\ref{polarization}) to (\ref{energy-density}), we see that electric,
magnetic, and magnetoelectric parts of energy combine to
\begin{align}
\label{energy1}
w = \frac12 \left(\alpha M_0^2 - \chi_e \gamma^2 M_0^4 \right) (\nabla\phi)^2 -\nonumber \\
- \chi_e \gamma M_0^2 \left(\partial_x \phi E_y - \partial_y \phi E_x\right) -
\frac{1}{2} \chi_e E^2.
\end{align}

Assume, for a moment, that ${\mathbf E} = 0$.
In this case expression for energy $W=\int w dV$ is similar to one of the XY model $W_{XY} = \frac12 \rho_s \int (\nabla\phi)^2 dS$
with an effective spin-wave stiffness
\begin{equation}
\label{rho}
\rho_s = (\alpha M_0^2 - \chi_e \gamma^2 M_0^4) h,
\end{equation}
where $h$ is the film thickness. Hence, in the absence of the electric field, the magnetoelectric coupling just renormalizes the spin-stiffness of the XY model.
For typical values of parameters (see Sec.\ref{Sec_possible_experimental_observation} and especially text after formula (\ref{gammaI}) for details): $\alpha M_0^2$ is greater than $\chi_e\gamma^2 M_0^4$ by several orders of magnitude; therefore, $\rho_s$ remains almost unchanged by magnetoelectric interaction and further we use $\rho_s \approx \alpha M_0^2 h$.

Despite this, some interesting features appear. It is well known that in the XY and related models the magnetic vortices play the crucial role \cite{KT1}. Vortex placed at the origin has the form
\begin{equation}
\label{vortex}
\phi = k \arctan\frac{y}{x} + \phi_0,
\end{equation}
where $k$ is the winding number of the vortex, $\phi_0$ is an arbitrary angle.
It turns out that magnetoelectric coupling affects vortices.
From (\ref{polarization}) it follows that polarization of magnetic vortex is ${\mathbf P} = -k \gamma\chi_e M_0^2 {\mathbf r}/r^2$.
This leads to the fact that a magnetic vortex acquires an electric charge $k q_e$ proportional to the winding number $k$ of the vortex\cite{Mostovoy:2006}, where
\begin{equation}
\label{qe}
q_e = 2\pi \gamma\chi_e M_0^2 h,
\end{equation}
\noindent where $\lambda_e$ is a vortex charge per unit film thickness.


Consider now what happens in non-zero electric field.
The $\phi$-dependent part of energy can be rewritten with the accuracy $O(\gamma)$ in the coupling constant as
\begin{align}
\label{energy2}
& W = \frac{h}{2} \int \left( \alpha M_0^2 (\nabla\phi)^2 -
2\chi_e \gamma M_0^2 \left(\partial_x \phi E_y - \partial_y \phi E_x\right) \right) dS = \nonumber\\
&= \frac{h \alpha M_0^2}{2} \int \left( (\partial_x\phi - \frac{\chi_e\gamma}{\alpha} E_y)^2 + (\partial_y\phi + \frac{\chi_e\gamma}{\alpha} E_x)^2 \right) dS
\end{align}
From this we see that the ground state in the electric field is reached when $\nabla\phi = (E_y, -E_x) \chi_e \gamma/\alpha$.

For example, a weak constant in-plane electric field modifies the ground state configuration into a cycloidal magnetic structure with a period $\lambda = 2 \pi \rho_s/\chi_e \gamma E$ and the wave-vector  perpendicular to the electric field \cite{Most1}. At $T=0$ this magnetic structure creates polarization ${\mathbf P} = \chi_e^2 \gamma^2 M_0^2 {\mathbf E}/\rho_s$ and gives contribution to electric susceptibility: $\chi_{cycloid} = \chi_e^2 \gamma^2 M_0^2/\rho_s$. We also see, that $\lambda \rightarrow \infty$ as $E \rightarrow 0$, and hence, finite-size magnetic orientational defects are not influenced by infinitesimal electric field.

When the electric field is not constant, nor weak, it is convenient to write it with the use of the electric field potential ${\mathbf E} = -\nabla\varphi$, so the coupling term (\ref{w_me}) can be rewritten as
\begin{align}
\label{Wme1}
W_{me} = h \chi_e \gamma M_0^2\int  \left(\partial_x\phi \, \partial_y\varphi - \partial_y\phi \, \partial_x\varphi \right) dS
\end{align}
Excluding singular points from the integration domain and making cuts in order that it was simply connected, we can integrate by parts and get:
\begin{align}
\label{Wme2}
W_{me} = h \chi_e \gamma M_0^2 \left( \oint \left( \partial_x\phi \hat{n}_y - \partial_y\phi \hat{n}_x \right)\varphi \, dl - \right. \nonumber \\
\left. -\int  \left(\partial_y\partial_x\phi - \partial_x\partial_y\phi \right)\varphi \, dS \right),
\end{align}
where the boundary integral over the edge of the sample $dl$ contains a normal to it, $\hat{n}$. Since the integration domain is simply connected and does not contain singularities, the second integral in (\ref{Wme2}) vanishes.

The first integral can be transformed using the fact that $d{\mathbf l} \sim (-\hat{n}_y, \hat{n}_x)$, so $\oint \left( \partial_x\phi \hat{n}_y - \partial_y\phi \hat{n}_x \right)\varphi \, dl = - \oint \varphi \, \nabla\phi d{\mathbf l}$. Integration contour includes not only outer boundary of the sample, but also all singular points of $\nabla\phi$. Since integration around the singularities is clockwise, we take out $-1$ and make it counterclockwise; $\oint \nabla\phi d{\mathbf l}= 2\pi k_i$, where $k_i$ is the winding number (vorticity) of $i$-th singular point. Integrals along cuts give $0$, since $\partial_i \phi$ is single-valued (though $\phi$ itself is not). Therefore, we obtain
\begin{align}
\label{Wme3}
W_{me} = h \chi_e \gamma M_0^2 \left( \sum_{cores} 2\pi k_i \varphi({\mathbf r}_i) - \oint\displaylimits_{edge} \varphi \nabla\phi d{\mathbf l}\right)
\end{align}
The first term in (\ref{Wme3}) has a form $\sum{q_i\varphi({\mathbf r}_i)}$. This means, that similarly to the case of zero electric field (\ref{qe}), in non-zero electric field vortices still have an electric charge proportional to еру winding number of the vortex $q_i = k_i q_e = k_i \cdot 2\pi h \chi_e \gamma M_0^2$. Therefore, vortices carry positive charge $q_e$ and antivortices carry negative charge $-q_e$.

However, from (\ref{Wme3}) we see that when the external electric field is applied, boundary effects become important. Consider a single vortex, which has electric charge (\ref{qe}) due to the magnetoelectric interaction. From the total electroneutrality it follows that the boundary of the sample also becomes charged; it
acquires charge of the same magnitude but the opposite sign compared to the vortex.
In a longitudinal external field, the boundary charge effectively shields the charge of the vortex,
and the value of screening depends on the geometry of the sample.
In this work we assume for simplicity that our sample is a disk with radius $R$. For such a disk screening of longitudinal constant electric field reduces the effective charge of vortices exactly twice \cite{Most1} (see Appendix A for further details). For axially symmetric electric fields with potential $\varphi(r)$ the integral in (\ref{Wme3}) gives just $q_{edge} \varphi(R)$ with $q_{edge} = h \chi_e \gamma M_0^2 \oint \nabla\phi d{\mathbf l}$. So for magnetically neutral systems (when $\oint \nabla\phi d{\mathbf l} = 0$) the edge charge is zero and (\ref{Wme3}) transforms to
\begin{align}
W_{me} = - q_e \sum_i k_i \varphi({\mathbf r}_i).
\label{Wme4}
\end{align}

Coupling energy (\ref{Wme3}) is additive both in $\phi$ and in $\varphi$. Additivity in $\phi$ tells us that magnetoelectric coupling does not influence the vortex-vortex magnetic interaction, which comes only from non-additive term $(\nabla\phi)^2$ in (\ref{energy1}). Therefore vortices interact as in XY model via $2D$ Coulomb potential \cite{KT1} $W = \pm 2q_m^2 \ln (r/a)$, where $a$ is the short-distance cut-off of order of the lattice spacing and $q_m$ is the ``magnetic charge'':
\begin{align}
q_m^2 = \pi\rho_s \approx \alpha M_0^2 h.
\label{qm}
\end{align}
Here and further we put for simplicity vorticities to be $k=\pm 1$, since vortices with higher winding numbers are unstable towards decomposition to these elementary ones.

The magnetic energy of a magnetically neutral system of $N$ vortices and $N$ antivortices can be written as\cite{Chaikin_Lubensky}:
\begin{align}
&W_m = - 2q_m^2 \sum_{i<j} k_i k_j \ln\frac{r_{ij}}{a}.
\label{Wm}
\end{align}
here $r_{ij}$ is the distance between $i$-th and $j$-th vortex cores.

For a sample with the number $N_1$ of vortices not equal to the number $N_2$ of antivortices formula (\ref{Wm}) can be generalized to %
\begin{align}
W_m = (N_1-N_2)^2 q_m^2 \ln \frac{R}{a} - 2q_m^2 \sum_{n<m} k_i k_j \ln\frac{r_{ij}}{a}.
\label{Wm_small}
\end{align}

We would like to emphasize that there is also purely electrostatic Coulomb interaction between the vortex cores, but it gives contribution $\sim \gamma^2$, which is negligibly small compared to the interaction $\sim\gamma$ with the external electric field. Therefore, in  what follows we adopt an approximation, in which a system of vortices and antivortices interact with each other only via the magnetic subsystem, and in addition,  they interact with the external electric fields due to the magnetoelectric coupling.


\section{Breaking the magnetic vacuum with electric field: vortex - antivortex pairs creation}

In this section we explore the basic idea of the paper, that a strong non-uniform electric field can break the ``magnetic vacuum''  of magnetoelectric material via creation of vortex-antivortex pairs.

\subsection{Geometry of the system}

\begin{figure}[tbh]
\centering
  \includegraphics[width=1.0\linewidth]{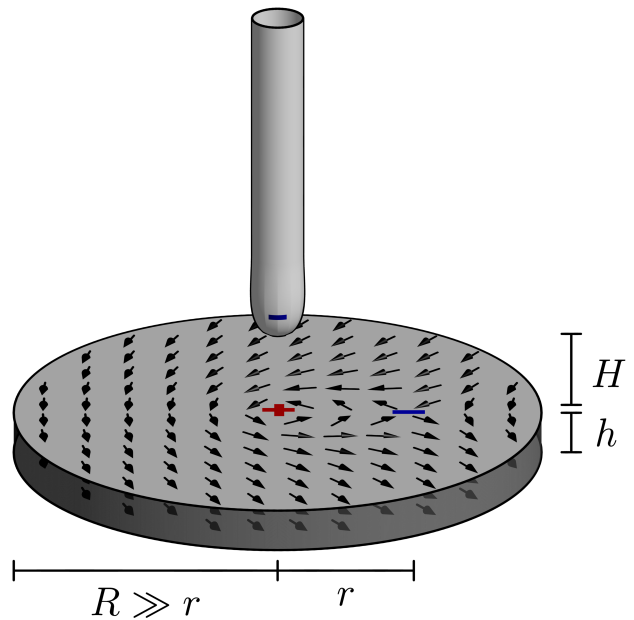}%
\caption{(Color online) Geometry of the system. Sample is a disk with radius $R$ and thickness $h$. Above its center at the height $H$ the tip of the cantilever with a negative charge $-Q_{tip}$ is placed.}
\label{fig_geometry}
\end{figure}

Fig.\ref{fig_geometry} shows typical experimental setup that we keep in mind (which was proposed in Refs. [\onlinecite{Pyatakov:2012}],[\onlinecite{Pyatakov:2010}]). An electric field is produced by a tip of a cantilever with a fixed point charge $-Q_{tip}<0$. Because of the magnetoelectric interaction, strong enough electric field makes possible creation of magnetic vortices. In this section we work in the limit of low temperatures, where there is no thermally activated vortex-antivortex (VA) pairs (for $T \ll \pi\rho_s = q_m^2$ their number is exponentially small).

Electric potential produced by the tip of the cantilever with charge $-Q_{tip}$ is
\begin{align}
\varphi(r) = -\frac{Q_{tip}}{\sqrt{r^2+H^2}}
\label{phi_tip}
\end{align}
where $H$ is the distance from the tip to the sample, $r$ is the polar radius of the point of observation.
We assume that the sample is thin: its thickness $h \ll H$, so we neglect the variation of the electric field in the transverse to the film direction inside the film.

\subsection{Critical tip charge for the creation of the first VA pair}

In this subsection we find critical value of electric field that creates the first vortex-antivortex pair at $T=0$ in a sample of big enough size to accommodate the pair as a whole ("big sample").

\begin{figure}[tbh]
\centering
  \includegraphics[width=1.0\linewidth]{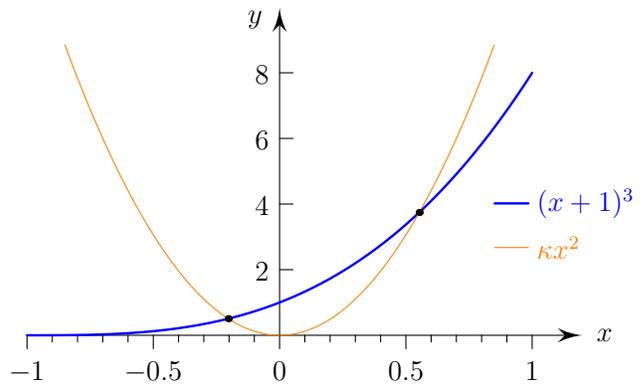}%
\caption{(Color online) Plots for LHS and RHS of Eq.(\ref{cubic}) for $\kappa = 3.5 > \kappa_0$. Two intersection points $x\approx -0.2$ and $x\approx 0.6$ are seen, while the third unseen intersection point $x\approx 8.9$ lies far to the right.}
\label{fig_cubic}
\end{figure}

Let the electric field produce one VA-pair with vortex situated at the center of the disk and antivortex placed at the distance $r$ from its center (in the considered case of big samples $r\ll R$). Energy of the pair as a function of $r$ consists of the magnetic energy (\ref{Wm}) of VA-interaction and magnetoelectric part (\ref{Wme4}):
\begin{align*}
W = 2 q_m^2 \ln\frac{r}{a} + Q_{tip} q_e \left( \frac{1}{\sqrt{r^2+H^2}} - \frac{1}{H} \right).
\end{align*}
Instead of $Q_{tip}$, $r$, and $H$ it is convenient to introduce dimensionless variables $\kappa$, $x$, and $h$:
\begin{align}
\label{dimensionless}
& \kappa = \frac{Q_{tip} q_e}{2 q_m^2 H}, \\
& x = \frac{r^2}{H^2}, \\
& h = \frac{H}{a}.
\end{align}
The total dimensionless energy expressed with the parameters $\kappa$, $x$, and $h$ is
\begin{align*}
W = \kappa \left( \frac{1}{\sqrt{x+1}} -1 \right) + \frac{1}{2} \ln x + \ln h
\end{align*}
Then, minimizing the energy with respect to the distance between the vortex and the antivortex $r$  by the condition $dW/dx = 0$
we obtain the following cubic equation
\begin{align}
\label{cubic}
(x+1)^3 = \kappa^2 x^2.
\end{align}
This equation always has a negative root (Fig.\ref{fig_cubic}). When parameter $\kappa$ increases from $0$ to some critical value $\kappa_0$ the equation also acquires a positive multiple root. This happens when the discriminant of the cubic equation is equal to $0$, from which we find $\kappa_0^2 = 3\sqrt{3}/2$ and $x_{min} = 2$, which means that for this critical value of charge the most favourable distance for VA-pair is $r_{min} = H\sqrt{2}$. For $\kappa > \kappa_0$ the potential energy always possesses a local minimum. Whether it will or will not be an absolute one depends on the dimensionless parameter $h = H/a \gg 1$.

The vortex-free state has zero energy. Then, the local minimum state with one VA pair becomes the global one when its energy becomes negative. This condition $W<0$ is fulfilled when:
\begin{align}
\frac{\kappa}{\sqrt{x_0+1}} < \kappa - \ln h - \frac{1}{2} \ln x_0
\label{energy_inequality}
\end{align}
where $x_0(\kappa)$ is the greatest root of (\ref{cubic}) that corresponds to the local minimum of energy; $x_0(\kappa) \geq 2$. Since the left hand side of (\ref{energy_inequality}) is positive, the above inequality can be fulfilled only if the right hand side is also positive, therefore, at least $\kappa > \ln h \gg 1$. For such big $\kappa$ equation (\ref{cubic}) always has 2 positive roots, with the greater one $x_0 \approx \kappa^2$ (and the lesser one $x \approx 1/\kappa$ corresponding to the local maximum of energy). Returning back to the original variables we find the optimal distance between the vortex and the antivortex that minimizes the energy $W$:
$$
r_0\equiv H\sqrt{x_0} \approx \frac{Q_{tip} q_e}{2q_m^2}.
$$
Substituting $x_0\approx \kappa^2$ in (\ref{energy_inequality}) with the accuracy $O(\kappa^{-1})$ we get
\begin{align}
\kappa - \ln h - \ln\kappa >  1
\label{energy_inequality2}
\end{align}
An approximate solution of the equation $\kappa - \ln h - \ln\kappa = 1$ is $\kappa \approx \ln h + 1 + \ln(\ln h+1) \approx \ln h$. From this we deduce the critical charge $Q_{tip}^{crit}$, above which the first VA-pair forms
\begin{align}
& \kappa_{crit} =  \frac{Q_{tip}^{crit} q_e}{2 q_m^2 H}  \approx \ln h\equiv\ln \frac{H}{a}; \\
& Q_{tip}^{crit}  = \frac{2 q_m^2 H}{q_e} \ln \frac{H}{a}.
\label{crit_charge}
\end{align}

For $Q_{tip} \gtrsim Q_{tip}^{crit}$ vortex-antivortex solution corresponds not only to a local but also to the global minimum of energy.

We also define the critical voltage of the tip as voltage produced at the nearest to the tip point of the sample:
\begin{align}
\varphi_0^{crit} = \frac{Q_{tip}^{crit}}{H} \label{vcrit} = \frac{2 q_m^2}{q_e} \ln \frac{H}{a}
\end{align}
For the same typical values of parameters as we used in Sec.\ref{Sec_Model}, we get $2 q_m^2/q_e \simeq 10..100 \mbox{V}$, so $\varphi_0^{crit} \simeq (10..100 \mbox{V}) \ln(H/a)$.
Higher values of tip voltages can lead to systems with several vortex-antivortex pairs.

\subsection{Vortex creation in small samples}

Above we considered the case of a big sample, when VA-pair lies far from its boundary. The opposite case of small samples is also possible and some estimates were made in Ref. [\onlinecite{Pyatakov:2010}]. If $r_0$ becomes greater than $R$ (or $R < Q_{tip} q_e/2q_m^2$), the antivortex leaves the sample, and because the positively  electrically charged vortex remains in its center, the boundary of the sample acquires a negative charge (see Appendix A for details). The energy of the system is
\begin{align*}
W = q_m^2 \ln\frac{R}{a} - q_e \Delta \varphi,
\end{align*}
where $\Delta\varphi = Q_{tip} (1/H-1/\sqrt{R^2+H^2})$ is the electric potential difference between the boundary and the center of the sample.
In this case the single vortex is created when energy $W$ becomes negative, i.e. when voltage
$$
\Delta\varphi = \frac{q_m^2}{q_e} \ln\frac{R}{a}
$$
is applied.

\section{Vortex distribution in continuous approximation}

In this section we construct an analytical theory in the continuous approximation for vortex distribution. It proves that even such a crude approximation captures many effects and allows us to find the vortex distribution profile, the number of vortices for given $Q_{tip}$, and the ground state energy of the system. These results are numerically verified in Sec.\ref{Sec_Numerical}.

\subsection{Self-consistent calculation of the vortex density $n(r)$ and the number of vortices $N$}
\label{sec_n_r}

The number of vortices in the system is not fixed (as if it were an external parameter), but rather is determined by the condition of the energy minimum. In this section we calculate self-consistently the number of vortices $N$ created by a charge $-Q_{tip}$ at $T=0$.

Consider the continuous limit (which means that at least $N \gg 1$, so $Q_{tip} \gg Q_{tip}^{crit}$, see (\ref{crit_charge})). Let $n_v({\mathbf r})$ and $n_a({\mathbf r})$ be concentrations of vortices and antivortices respectively. Let $n({\mathbf r}) = n_v({\mathbf r}) - n_a({\mathbf r})$. At $T=0$, regions of non-vanishing $n_v$ and $n_a$ does not intersect, since if they did, the vortices in the intersection area would annihilate with antivortices in order to lower the energy of the system.

Consider a ring between radii $r$ and $r+dr$. In the ground state there is a balance between the electric and magnetic interactions. Magnetic interaction pushes vortices outside while external electric force pulls them inside the ring (and vice versa for antivortices). From (\ref{phi_tip}) external electric force acting upon a given area element with electric charge $q_e dN = q_e n dS$ is:
\begin{align}
\label{Fel}
F_{el} = q_e dN \frac{d\varphi}{dr} =  Q_{tip} q_e n dS \frac{r}{(r^2+H^2)^{3/2}}.
\end{align}
From (\ref{Wm}) magnetic force is $2 q_m^2/r$; we can now calculate magnetic force acting at the same area element with magnetic charge
$dq_m = q_m n dS$, using the Gauss's theorem from electrostatics: an axially-symmetric ring acts on the outer charges with such a force as if all charges of the ring were concentrated in its center and does not act on the inner charges, therefore the magnetic force is
\begin{align}
\label{Fmagn}
F_{magn} = \frac{2 dq_m q_m^{inside}}{r},
\end{align}
where $q_m^{inside} = q_m \int n({\mathbf r}) d^2 r = 2\pi q_m \int_0^r n(r) r dr$ is a net magnetic charge inside the considered ring.

\begin{figure}[tbh]
\centering
  \includegraphics[width=0.9\linewidth]{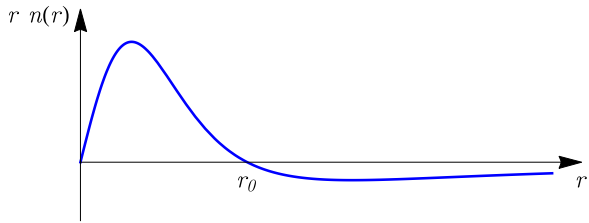}%
\caption{(Color online) Schematic plot for $r n(r)$ vs $r$. Here $r n(r)$ is proportional to the number of vortices at distance $r$ from the disk center; $n(r)$ is given by (\ref{n_r}).}
\label{plot_n_r}
\end{figure}

Equating (\ref{Fel}) and (\ref{Fmagn}) we get:
\begin{align}
\label{int_n_r}
\frac{Q_{tip} q_e r}{(r^2+H^2)^{3/2}} = \frac{4\pi q_m^2}{r} \int_0^r n(r) r dr
\end{align}
Multiplying by $r$ and differentiating this with respect to $r$ we obtain the vortex distribution density profile:
\begin{align}
\label{n_r}
n(r) = \frac{Q_{tip} q_e}{4\pi q_m^2} \frac{2H^2-r^2}{(r^2+H^2)^{5/2}}
\end{align}
The $r$-dependence of $r n(r)$ that is proportional to the number of vortices at distance $r$ from the disk center is shown in Fig.\ref{plot_n_r}. Vortices are concentrated inside a small circle of radius $r_0 = H\sqrt{2}$, antivortices are smeared over the sample outside $r_0$ with their density decaying as $r n(r) \sim -1/r^2$.

From (\ref{int_n_r}) we get the total topological charge of vortices and antivortices inside a circle of radius $r$:
\begin{align}
\label{Nr}
N(r) = 2\pi\int_0^{r} n(r) r dr = \frac{Q_{tip} q_e}{2 q_m^2} \frac{r^2}{(r^2+H^2)^{3/2}}
\end{align}

Now we can calculate the total number of vortices (which for big samples is the same as the number of antivortices). The border between vortices and antivortices lies at $r_0 = H \sqrt{2}$. Using this we get
\begin{align}
\label{Nselfconsistent}
N = \frac{1}{3^{3/2}}\frac{Q_{tip} q_e/H}{q_m^2}
\end{align}


For a small system with radius $R$ the number of antivortices can be less than the number of vortices, and there will be a linear charge density at the edge of the sample. This edge charge is
\begin{align*}
q_{edge} = - q_e N(R) = -\frac{Q_{tip} q^2_e}{2 q_m^2} \frac{R^2}{(R^2+H^2)^{3/2}}
\end{align*}

\subsection{Total ground state energy of the system}

Here we calculate the magnetic and electric energy of system of vortices and antivortices in the ground state.

First, we calculate the magnetic energy. 
We use the continuous version of (\ref{Wm}), which can be obtained by the substitution $k_i \rightarrow n({\mathbf r}) dS \equiv n(r) dS$
\begin{align*}
W_m = 
-2q_m^2 \int d^2 r \, n(r) \int\displaylimits_{r' < r} d^2 r' n(r') \ln\frac{|{\mathbf r} - {\mathbf r}'|}{a}
\end{align*}
Integral over $d^2 r'$ can be calculated using Gauss's theorem for the potential energy:
$\int_{r'<r} d^2 r' n(r') \ln |{\mathbf r} - {\mathbf r}'|/a = \ln (r/a) \int d^2 r' n(r') \equiv \ln (r/a) N(r)$, where $N(r)$ is given by (\ref{Nr}), so we obtain
\begin{align*}
W_m = - \frac{Q_{tip}^2 q_e^2}{2q_m^2} \int\displaylimits_0^{\infty} dr \, \frac{r^3 (2H^2-r^2)}{(r^2+H^2)^4} \ln\frac{r}{a}
\end{align*}
Integrating this we get:
\begin{align}
\label{ground_Wm}
W_m = \frac{Q_{tip}^2 q_e^2}{16 q_m^2 H^2}
\end{align}

Second, we find the energy of the magnetic vortices in external electric field (\ref{phi_tip}):
\begin{align}
\label{ground_Wel}
W_{me} =  q_e \int_0^{\infty} d^2 r \, n(r) \varphi(r) = -\frac{Q_{tip}^2 q_e^2}{8 q_m H^2}
\end{align}

Summing (\ref{ground_Wm}) and (\ref{ground_Wel}) we get the total ground state energy of the system in the continuous limit:
\begin{align}
\label{ground_W}
W = -\frac{Q_{tip}^2 q_e^2}{16 q_m^2 H^2}.
\end{align}
Since it is negative, and the energy of uniform vortex-free state is taken to be $0$, then in the continuous limit formation of vortex structures is always energetically favourable. This estimate, as well as the continuous limit itself, is valid only for $Q_{tip} \gg Q_{tip}^{crit}$.

Concluding this section, we note that despite continuous limit being a crude approximation, it gives us the distribution density of induced vortices and their number, which are numerically verified in Sec.\ref{Sec_Numerical}. The next section addresses some effects that can not be captured in the continuous approximation.





%

\section{Calculation of polarizability: discreteness effects}
\label{Sec_polarizability_RMT}

In this section we find the polarizability of big samples with zero total magnetic vorticity, for which even the most distant from the tip antivortices do not feel the boundaries of the sample. As explained below, continuum approach cannot be applied to polarizability calculation and we have to deal explicitly with the effects of discreteness of vortex distribution over the sample.

We study two limiting cases: (A) the case of very low temperatures $T\ll q_m$ when only tip-induced vortices are present; (B) the case of higher temperatures $T\lesssim q^2_m/2 \simeq T_{BKT}$, when close to the Berezinskii-Kosterlitz-Thouless temperature there also exist some amount of thermodynamically activated pairs.

\subsection{$T \ll q_m^2$}

Consider a system of $N$ vortices and $N$ antivortices. As a first approximation we can neglect the effect of the finite size of the positive vortex ``nuclei'' and consider it as a point-like one, neglecting its contribution to the polarizability. Then, remaining $2D$ system of $N$ antivortices has $2N$ degrees of freedom, therefore, $2N$ normal modes.

Consider the system near its local energy minimum. Polarizability at $T=0$ is largely determined by the most soft mode of the system. If the system possesses a zero mode that couples with the electric field, then an arbitrary small perturbation leads to huge distortions, so polarizability of such state is infinite.
However, temperature fluctuations can effectively smear the low lying modes, which makes the polarizability finite and strongly temperature dependent.

The problem of finding polarizability can be exactly analytically solved for small fixed values of $N$, but the exact solution seems to be inaccessible for arbitrary large $N$. Here we employ the strategy similar to ``energy landscape approach'' used for overcooled liquids, glasses and spin glasses, proteins folding, and melting of small clusters \cite{Wales:book}.

Let $r_i = r_i^{(0)} + \Delta r_i$ ($i=1$,...,$2N$), where $\Delta r_i$ are deviations of cartesian coordinates of $N$ antivortices from the local minimum, with the first $N$ components giving us $x$-coordinates of antivortices, and the second $N$ components giving their $y$-coordinates: $r_1 = x_1$, ..., $r_N = x_N$; $r_{N+1} = y_1$, ..., $r_{2N} = y_N$.
The total energy $W = W^{(0)} + W^{(1)}$, where $W^{(0)} = \sum_{i,j} V(x_i, y_i; x_j, y_j)+\sum_{i} V_{tip}(x_i, y_i)$ is the sum of magnetic energy of antivortices and electrostatic energy interaction of antivortices with the tip. Small perturbation
$W^{(1)} = q_e E \sum_{i=1}^N \Delta r_i$ is the energy of interaction with the infinitesimal in-plane electric field $E$ in the $x$-direction.

We employ the Hessian matrix:
\begin{align}
K_{i,j} = \frac{\partial^2 W^{(0)}}{\partial r_i \partial r_j}
\end{align}
Near the local minima, energy $W^{(0)}$ can be written as a positive definite quadratic form (vector $\Delta{\mathbf r} = (\Delta r_1,...,\Delta r_{2N})$):
\begin{align}
W^{(0)} = \frac{1}{2} \sum_{i,j=1}^{2N} \Delta r_i K_{ij} \Delta r_j = \frac{1}{2} \Delta{\mathbf r} K \Delta{\mathbf r}
\end{align}

Symmetric matrix $K_{ij}$ can be diagonalized by some orthogonal transformation $O$: $\tilde{K} = O K O^{-1}$, with coordinates $\Delta r_i$ being accordingly transformed to normal coordinates $R_i$: $\Delta{\mathbf r} = O{\mathbf R}$.
Then $W^{(0)}$ transforms as
\begin{align}
W^{(0)} = \frac{1}{2} {\mathbf R} \tilde{K}  {\mathbf R}
\end{align}
where $\tilde{K} = diag\{K_1, ..., K_{2N}\}$, $K_i$ are eigenvalues of the Hessian matrix, which play an important role as we will see below.

Since $\Delta r_i = (O {\mathbf R})_i = \sum_{j=1}^{2N} O_{ij} R_j$, $W^{(1)}$ transforms as
\begin{align}
W^{(1)} = q_e E \sum_{i=1}^N \Delta r_i = q_e \sum_{j=1}^{2N} o_j R_j
\end{align}
where
\begin{align}
o_j = \sum_{i=1}^N  O_{ij}.
\end{align}

The partition function can be written in terms of normal coordinates $R_i$ as
\begin{align}
\label{Z_E}
& Z(E) = \int \prod_{i=1}^{2N} d\Delta r_i \exp\left( -\beta(W^{(0)}+W^{(1)}) \right) = \nonumber \\
&= \int_{-L}^L \prod_{i=1}^{2N} dR_i \exp\left( -\frac{1}{2} \beta \sum_{j=1}^{2N} R_j K_{j} R_j - \beta q_e E\sum_{j=1}^{2N} o_j R_j\right)
\end{align}
Here we introduced a cutoff $L$, which takes into account the fact that for big enough $R_i$ potential $W^{(0)}$ is no longer a parabolic one. $L$ corresponds to such displacement of antivortices that shifts the point in the configuration space from the local minimum to the nearest saddle point;
we expect $L$ to be of order of distance between nearest antivortices.

Polarization of antivortices with negative charges $-q_e$ is
\begin{align}
p_x(E) = -q_e \sum_{i=1}^N \langle r_i \rangle = \frac{T}{Z(E)} \frac{\partial Z}{\partial E}
\end{align}
Polarizability can be found as
\begin{align}
\alpha = \frac{\partial p_x}{\partial E} \bigg|_{E=0} =
\frac{T}{Z(0)} \frac{\partial^2 Z}{\partial E^2}\bigg|_{E=0}
\end{align}
Substituting here the partition function (\ref{Z_E}) we get
\begin{align}
\alpha =& \frac{q_e^2}{T} \, \sum_{i=1}^{2N} o_i^2 \, \frac{\displaystyle\int_{-L}^L dR_i R^2_i \exp\left( -\frac{1}{2} \beta K_i R^2_i\right)}
{\displaystyle\int_{-L}^L dR_i \exp\left( -\frac{1}{2} \beta K_{i} R^2_i \right)}
\label{alpha_from_Z}
\end{align}
%

As $T\rightarrow 0$ ($\beta \rightarrow \infty$) Gaussian integrals converge very fast and cut-off $L$ does not play any role: we can extend the limits of integration up to infinity.
Then from (\ref{alpha_from_Z}) we find
\begin{align}
\alpha(T\rightarrow 0) = \frac{q_e^2}{T} \sum_{i=1}^{2N} o_i^2 \frac{1}{\beta K_i} = q_e^2 \sum_{i=1}^{2N} \frac{o_i^2}{K_i}
\end{align}
From this we see that in the limit $T\rightarrow 0$ polarizability $\alpha$ is determined by the most ``soft'' modes $K_i$.

\begin{figure}[tbh]
\centering
  \includegraphics[width=0.9\linewidth]{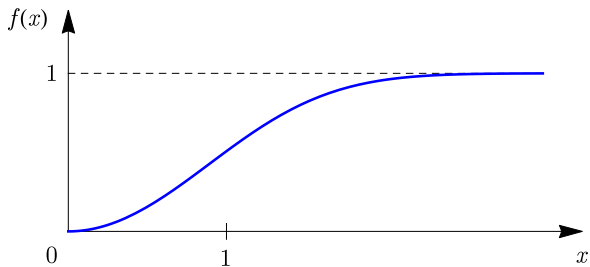}%
\caption{(Color online) Plot of the cut-off function $f(x)$ defined by formula (\ref{fx}). As $x\rightarrow0 $, $f(x)\sim \frac{2}{3}x^2$; as $x\rightarrow +\infty$, $f(x)\sim 1-\frac{2}{\sqrt{\pi}}x e^{-x^2}$.}
\label{fig_fx}
\end{figure}

Now consider the case of non-zero $T$. Performing integrations in (\ref{alpha_from_Z}) we see that cut-off $L$ now enters the expressions for polarizability explicitly:
\begin{align}
\alpha(T)  = q_e^2 \sum_{i=1}^{2N} \frac{o_i^2}{K_i}  f\left(\sqrt{\frac{\beta K_i}{2}}L\right),
\label{alpha_T}
\end{align}
where we denoted
\begin{align}
f(x) = 1-\frac{2xe^{-x^2}}{\sqrt{\pi} \erf x},
\label{fx}
\end{align}
playing the role of smooth cut-off function for the terms in the sum (\ref{alpha_T}) with the lowest eigenvalues $K_i$.
Function $f(x)$ is sketched in Fig.\ref{fig_fx}.

In the limit of big number of vortices and antivortices $N \gg 1$ the sum (\ref{alpha_T}) can be approximated by an integral:
\begin{align}
\alpha(T)  \approx q_e^2 \int_{K_{min}}^{K_{max}} \frac{o^2(K)}{K}  f\left(\sqrt{\frac{\beta K}{2}}L\right) D(K) dK
\label{alpha_integral}
\end{align}
where $D(K)$ is the density of $K$-modes.

In order to make some estimates of $\alpha(T)$ and deduce its leading temperature dependence we use the following three approximations.

(i) We estimate $D(K)$ using random matrix theory for an Euclidean ensemble \cite{Parisi:1999}. Amir et.al. \cite{Amir:2010} analytically calculated the density of eigenvalues for certain 1D dilute systems with short range potential and obtained $D(K) \sim 1/K$; for higher dimensions they predicted that there should be some correction
\begin{align}
D(K) \sim 1/K^{1-\eta},
\label{D_K}
\end{align}
where $\eta > 0$ depends on the dimensionality of the system and on the form of the particle interaction potential.
In Sec.\ref{Sec_big_samples_numerical} we check this assumption by exact numerical diagonalization of the Hessian matrix and find the exponent $\eta$.

(ii) We replace the smooth cut-off function with a sharp Heaviside theta function $f(x) \rightarrow \theta(1-x)$, so the lower limit of integration becomes $T/2L^2$. This can be done since $f(x) \sim \frac{2}{3} x^2$ for $x \rightarrow 0$ and integral (\ref{alpha_integral}) converges at the lower limit as $\int K^{1-\eta} dK$.

(iii) Since coupling to the electric filed is determined only by $x \rightarrow -x$ symmetry of the mode, we assume that all modes couple to the electric field with approximately the same strength, apart from fluctuations on small-$K$ scale:
\begin{align}
\langle o^2 (K) \rangle = \frac{1}{\Delta K} \int_K^{K+\Delta K} o^2 (K) dK \approx \const (K)
\end{align}
For example, if the stationary configuration $r^{(0)}$ is symmetric under reflection $x\rightarrow -x$, then all modes are two-fold degenerate and either symmetric (does not couple to the electric field) or antisymmetric (couple to the electric field with the same weight).

Finally, using the above three approximations, we obtain the following temperature dependence of polarizability:
\begin{align}
\alpha(T) \sim \int_{T/2L^2}^{+\infty} \frac{D(K) dK}{K} \sim \int_{T/2L^2}^{+\infty} \frac{dK}{K^{2-\eta}} \sim
\frac{1}{T^{1-\eta}}
\label{alpha-1sqrtT}
\end{align}
Since the integral converges we extended the upper limit of integration up to $\infty$. This estimate holds as long as $T \gg K_{min} L^2$. For very low temperatures $T \ll K_{min} L^2$, $\alpha \sim 1/K_{min}^{1-\eta} = const(T)$.

The exponent $\eta$ is in a sense a non-universal one and it should depend not only on the dimensionality of the system, but also on the form of the particle interaction and external potentials.
We numerically test the validity of the above approximations and the $\alpha\sim 1/T^{1-\eta}$ dependence for our particular case in Sec. \ref{Sec_big_samples_numerical}.

\subsection{Some estimates for $T \lesssim q_m^2/2$}

Now consider the case $T \lesssim q_m^2/2$. Then there also exist VA pairs created by thermal fluctuations.
The profile of the total vorticity $n({\mathbf r}) = n_v({\mathbf r}) - n_a({\mathbf r})$ remains the same as it was found in \ref{sec_n_r} from the force balance considerations (formula (\ref{n_r}) and Fig.\ref{plot_n_r}). The difference is that now the regions of non-vanishing $n_v(\mathbf{r})$ and $n_a(\mathbf{r})$ intersect with each other. This describes the process of penetration of antivortices in the vortex region and vice versa because of VA pairs formation.


Now there will be two contributions to polarizability:

(i) contribution of tip-induced vortices, but now their interaction will be renormalized by the Kosterlitz-Thouless dielectric function: $W(r)= \pm2q_m^2 \ln(r/a) / \epsilon(r)$;

(ii) pure contribution of thermally activated dipole pairs away from the tip.

\noindent Below we consider only the renormalized contribution (i) (pure contribution of thermally activated vortices (ii) is considered in Appendix B).



Renormalized interaction energy is $W(r)= \pm2q_m^2 \ln(r/a) / \epsilon_{KT}(r)$. As long as $\chi_{KT} = (\epsilon_{KT}-1)/2\pi \lesssim 1$, which holds in almost the whole range $T < T_{BKT} \approx q_m^2/2$ except in a small temperature interval in the vicinity of $T_{BKT}$, there will be no qualitative influence of the thermal VA-pairs onto the field induced vortices.

We can estimate the melting temperature for the induced ``magnetic atom'' patterns. The pattern starts to melt at a temperature $T_c$ when the position fluctuations become comparable to intervortex distance $\Delta r \sim 1/\sqrt{n_v}$ ($n_v$ is a local vortex concentration).
Make an estimate for the vortex situated in the center of the disk. Here only electric potential influences, which is almost parabolic:
\begin{align*}
W_e(r) \approx \frac{k r^2}{2}, \qquad k = \frac{2 Q_{tip} q_e}{H^3}
\end{align*}
Using (\ref{Nselfconsistent}) and $n_v \sim \pi r_0^2/N = 2\pi H^2/N$ we get:
%
\begin{align*}
T_c \sim \frac{k(\Delta r)^2}{2} \sim \frac{k}{2 n_v} \sim \frac{k H^2}{2N} \sim \frac{2 Q_{tip} q_e H^2}{2 H^3 N} \sim q_m^2 \sim T_{BKT}.
\end{align*}
Hence, the magnetic pattern is stable up to $T \lesssim T_{BKT}$. For $T>T_{BKT}$ thermodynamically induced vortex pairs dissociate and break the local magnetic order.


\section{Numerical simulation}
\label{Sec_Numerical}

In this section we describe the procedure and the results of numerical simulation.

\subsection{Numerical procedure}

We modeled a system of point-like vortices and antivortices, which interact between themselves (with the magnetic energy (\ref{Wm})) and with the external electric field created by cantilever's tip (with energy (\ref{Wme4}), electrostatic potential is given by (\ref{phi_tip})).
The system shows ``glassy'' behavior, i.e. there are many local minima that are close to the global one.
In order to find a state corresponding to such a local energy minimum we performed simulated annealing using Monte Carlo (MC) method \cite{Kirkpatrick:1983}. Three types of MC-steps were used: movement of a vortex or an antivortex, creation of VA pair, and destruction of VA-pair.
We cooled the system from high temperatures $T_{start} \simeq q_m^2 > T_{BKT}\simeq q_m^2/2$ (which allowed the system to find the optimal number of vortices) down to as low as $T_{stop} = 10^{-6} q_m^2$, depending on the numerical experiment.
Geometrical cooling schedule was used \cite{Kirkpatrick:1991}:
\begin{align*}
T_k = T_{start} \gamma^{[k/N_{it}]}
\end{align*}
with $\gamma = 0.99$ and typically $N_{it} = 100 000$ number of Monte Carlo iterations at a given temperature.

The simulation was performed for the following ratios between the lattice constant $a$, distance $H$ from the tip to the sample, and radius $R$ of the sample: $a = 0.01 H$, $R = 100 H$.

\subsection{Number of vortices and the critical field}

\begin{figure}[tbh]
\centering
  \includegraphics[width=0.9\linewidth]{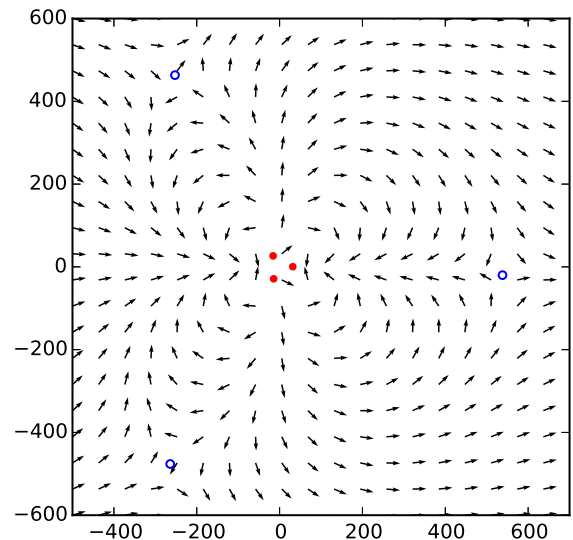}%
\caption{(Color online) Configurations of vortex-antivortex system for $Q_{tip} q_e/H= 23 q_m^2$ at $T=0.001 q_m^2$.
Vortex cores are (filled circles) and antivortex cores (open circles) are shown; arrows show magnetization. Coordinates are given in the units of lattice spacing $a$.}
\label{fig_M}
\end{figure}

\begin{figure}[tbh]
\centering
\subfloat[\label{a}]{%
  \includegraphics[width=0.5\linewidth]{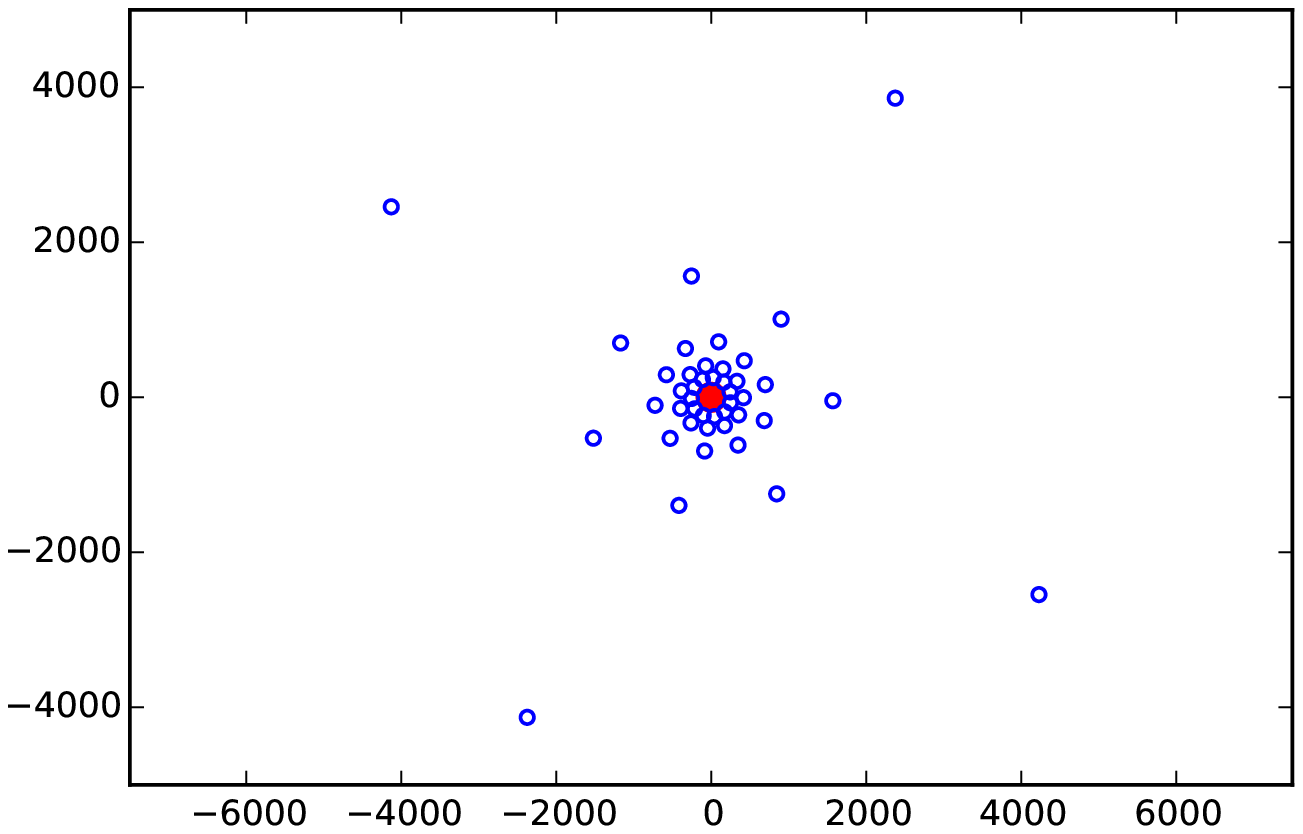}%
}
\subfloat[\label{b}]{%
  \includegraphics[width=0.5\linewidth]{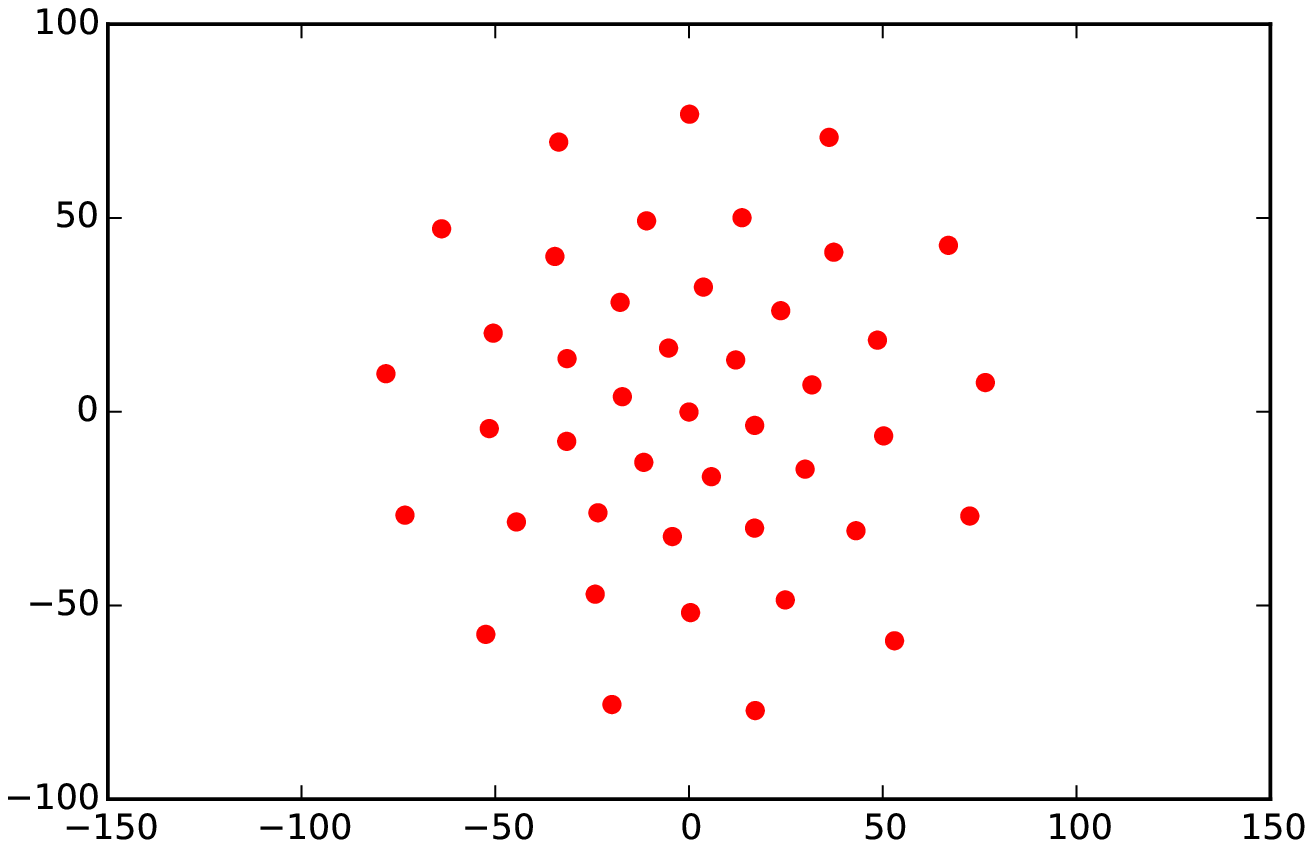}%
}
\caption{(Color online) Configurations of vortex-antivortex system for $Q_{tip} q_e/H= 240 q_m^2$ at $T= 10^{-6} q_m^2$.
Vortex cores are (filled circles) and antivortex cores (open circles) are shown. a) Antivortex subsystem, vortex ``nucleus'' cannot be resolved; b) vortex subsystem.}
\label{fig_numerical_N}
\end{figure}

\begin{figure}[tbh]
\centering
  \includegraphics[width=1.0\linewidth]{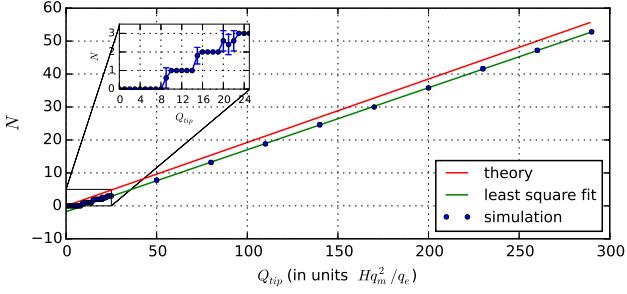}%
\caption{(Color online) Plot for $N$ vs $Q_{tip}$. The inset shows zoom of the plot for range $Q_{tip}=0\div24 H q_m^2/q_e$. From the inset the critical tip charge is $Q_{tip}^{crit} \approx 9 H q_m^2/q_e$.}
\label{fig_numerical_NvsQ_plot}
\end{figure}


As we described in detail in Sec.III and IV, charged tip produces magnetic patterns (``magnetic atoms'') in the system under consideration.
Fig.\ref{fig_M} shows typical low temperature ($T=0.001 q_m^2$) distribution of magnetization in the system for moderate values of tip voltages ($\varphi = Q_{tip}/H = 23 q_m^2/q_e$), for which $N=3$ vortices and an $3$ antivortices are created.

Fig.\ref{fig_numerical_N} shows typical low-temperature configuration of the system for higher values of tip's voltage ($Q_{tip} q_e/H= 240 q_m^2$). For the depicted configuration the observed number of vortices is $N^{num} = 43$, whereas analytical formula (\ref{Nselfconsistent}) gives $N^{analyt} = 240/3^{3/2} \approx 46.2$, this mismatch is explained below.
Vortices and antivortices are arranged in a highly regular structure that is reminiscent of the structure of an atom: negatively charged antivortex outer concentric circular ``shells'' surround positive vortex ``nucleus''. Fig.\ref{fig_numerical_N}b depicts only this vortex ``nucleus'' of the ``atom''; arrangement of the vortices, interacting via 2D Coulomb-like $\log r$ magnetic potential, is similar to one of the trapped single species Coulomb plasma, where charges are confined to a plane and interact via 3D Coulomb $1/r$ potential \cite{Bedanov:1994,Radzvilavicius:2011}. Similar patterns of of magnetic skyrmion systems were observed recently \cite{Zhao:2016}.

Fig.\ref{fig_numerical_NvsQ_plot} shows $N(Q_{tip})$ dependence for the ground states of the system with varying $Q_{tip}$.
We see that analytical result (\ref{Nselfconsistent}) is in a good agreement with the numerical modeling. Analytical theory, which was derived in a continuous approximation, gives the correct slope, but slightly overestimates the number of vortices.

We also see that the analytical result for the critical charge that creates the first VA-pair (\ref{crit_charge}) gives for the considered parameters ($a = 0.01 H$) $Q_{tip}^{crit} = 2\ln 100 q_m^2 H/q_e \approx 9.2 q_m^2 H/q_e$, whereas the numerical simulation gives $Q_{tip}^{crit} \approx 9 q_m^2 H/q_e$ (inset of Fig.\ref{fig_numerical_NvsQ_plot}). This mismatch $\Delta Q = Q_{tip}^{crit}$ between the continuous theory and simulation for $Q_{tip}$ at a given $N$ leads to mismatch
in $N$ for given $Q_{tip}$, and the average mismatch should be of order of $\Delta N \approx \Delta Q q_e/ 3^{3/2} q_m^2 H \approx 1.7$, moreover real step-like dependence gives an additional $\pm1$ deviation from the continuous formula (\ref{Nselfconsistent}), which explains the above mismatch $\Delta N \approx 3.2$ between the analytical formula and the numerical simulation for $Q_{tip} q_e/H= 240 q_m^2$.

\subsection{Polarizability of big samples}

\label{Sec_big_samples_numerical}

In this section we numerically find the temperature dependence of polarizability of big samples. In order to find polarizability we performed cooling at a fixed value of electric field $E$, then independently repeating this for different values of $E$. From the theoretical analysis of Sec.\ref{Sec_polarizability_RMT} we expect the polarizability to be strongly temperature-dependent.

\begin{figure}[tbh]
\centering
  \includegraphics[width=1.0\linewidth]{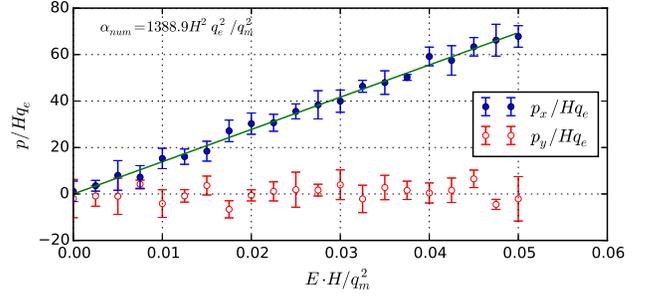}%
\caption{(Color online) Plot for dipole moment $p$ vs electric field $E$ at $T=0.001 q_m^2$ for big system (for $Q_{tip} q_e/H= 200 q_m^2$, $N=36$). Filled circles show $p_x$ dependence (component parallel to the electric field), open circles -- $p_y$ dependence (component perpendicular to the electric field).}
\label{fig_numerical_PvsE0_plot_big}
\end{figure}

Fig.\ref{fig_numerical_PvsE0_plot_big} shows the variation of the $x$-th and $y$-th components of the dipole moment with the applied electric field for fixed temperature $T=0.001 q_m^2$. Linear fit for $p_x(E)$ dependence gives polarizability $\alpha_{num} = 1388.9 H^2 q_e^2/q_m^2$.

\begin{figure}[tbh]
\centering
  \includegraphics[width=1.0\linewidth]{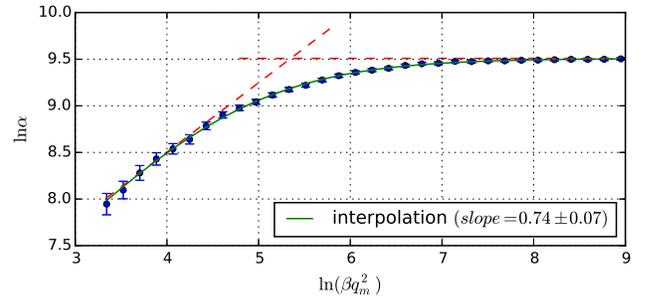}%
\caption{(Color online) $\ln\alpha$ vs $\ln(\beta q_m^2)$ plot (for $Q_{tip} q_e/H= 200 q_m^2$, $N=36$). Symbols show data with its standard deviation, solid line shows fitting. Slope of the dashed inclined line is $0.74 \pm 0.07$}
\label{fig_numerical_alpha_vs_beta}
\end{figure}

Fig.\ref{fig_numerical_alpha_vs_beta} shows the temperature dependence of polarizability in log-log coordinates, which in the limit of small $\beta$ is fitted by $\ln\alpha \approx 0.74 \ln(\beta q_m^2) + \const$. This dependence is consistent with the estimate (\ref{alpha-1sqrtT}), when we choose $\eta \approx 0.26$, which gives then $\alpha \sim 1/T^{0.74}$.

\begin{figure}[tbh]
\centering
  \includegraphics[width=1.0\linewidth]{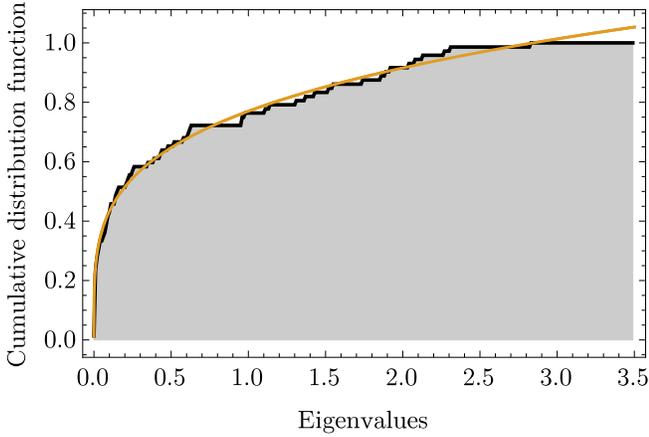}%
\caption{(Color online) Cumulative distribution function of eigenvalues $\int_0^K D(K) dK$ vs $K$. Thick broken line -- data from exact diagonalization of $72\times72$ Hessian matrix, thin curve -- best fit $\sim K^{0.25}$. This leads to $D(K) \sim K^{-0.75}$.}
\label{fig_eigenspectrum}
\end{figure}

Using the exact numerical diagonalization we have found an eigenspectrum (Fig.\ref{fig_eigenspectrum}), from which indeed exponent $\eta \approx 0.25$.

We propose that such non-trivial temperature dependence of polarizability of finite clusters is a generic one.
For glassy systems potential energy landscape of the system possesses many local minima that are close to the global one; adding one VA pair to the system completely changes this landscape and its local minima, but preserves its Hessian eigenvalue distribution. Then we argue that temperature fluctuations smear the fine details of this energy landscape, effectively cutting off the low-$K$ (low frequency) normal modes $K_i \ll T/L^2$. This makes polarizability of the system strongly temperature dependent with a non-trivial dependence $\alpha\sim 1/T^{1-\eta}$.
The exponent $\eta$ is non-universal one and depends on the density of Hessian eigenvalues. For example, for single-species 2D plasma confined by quadratic external potential, with particles interacting via 3D Coulomb $1/r$ potential $D(K) \sim 1/K^{0.33}$ [\onlinecite{Shweigert:1995}], so from (\ref{D_K}) in this case $\eta \approx 0.67$.

\section{Possible experimental observation}
\label{Sec_possible_experimental_observation}

In this section we consider the possibilities of experimental observation of the described effects and make estimates for the required electric fields.
The most plausible candidates possessing magnetoelectric effect seem to be multiferroic materials. However the most of the known multiferroics are antiferromagnets with the staggered magnetization playing the role of the order parameter. We believe that theoretical results presented in the previous sections are applicable to them at least qualitatively if not quantitatively. The important remark should be now kept in mind: by vortex/antivortex  we understand now $\pm 2\pi$ rotation of the staggered magnetization vector.
%
Also in order to ease the comparison with the experimental data we use SI units throughout this section; for transferring from Gaussian to SI units substitutions $\chi_e \rightarrow \chi_e \epsilon_0$, $Q_{tip} \rightarrow Q_{tip}/4\pi\epsilon_0$ should be made.


\subsection{Theoretical estimates for real materials}

First, we consider the potential materials where the predicted effects could be found. In order to observe creation of ``magnetic atoms'' by electric field in a thin film we need a fairly strong coupling between the electric and magnetic subsystems is needed.
Below we make estimates for type-I and type-II multiferroics.

\emph{Type-I multiferroics}. We consider prototypical type-I multiferroic -- $\mathrm{Bi Fe O_3}$. Its single crystals are rhombohedral at room temperature with the space group $R3c$ and pseudocubic unit cell with $a\approx \SI{4}{\angstrom}$ and G-type antiferromagnetic structure [\onlinecite{Lebeugle:2008PRL}], its spontaneous polarization $P \simeq \SI{1}{C/m^2}$ [\onlinecite{Lebeugle:2007APL},\onlinecite{Neaton:2005}], which induces cycloidal twisting of the staggered magnetisation with the period $\lambda \simeq 620\div \SI{640}{\angstrom}$ [\onlinecite{Sosnowska:1982},\onlinecite{Lebeugle:2008PRL}], dielectric constant $\epsilon = 1+\chi_e \simeq 50$ [\onlinecite{Lu:2010}],

In order to estimate constant $\alpha$ characterizing the strength of the exchange interaction (see eq. (\ref{w_m}) above), we use the exchange constant for nearest neighbor Fe-Fe interaction $J=\SI{10}{meV}$ [\onlinecite{Delaire:2012}]. Hence for the BCC pseudocubic setting we obtain exchange stiffness $A \equiv \alpha M_0^2/2 = J a^2/V = \SI{4}{pJ/m}$ which is consistent with the estimate deduced from the Neel temperature \cite{Sando:2013} ($V\approx a^3$ is the volume of the pseudocubic unit cell).

Now using (\ref{w_m}), (\ref{w_me}), (\ref{M}) and taking $\phi = k x$, where $k=2\pi/\lambda$, we get
\begin{align}
w_{me} + w_m= -\gamma P_y k M_0^2 + \frac{\alpha}{2} k^2 M_0^2
\label{w_m_spiral}
\end{align}
Minimizing this energy density with respect to $k$ we obtain $k=\gamma P_y/\alpha = 2\gamma P_y M_0^2/A$ or
\begin{align}
\gamma M_0^2 = k A/2P \equiv \pi A/\lambda P \simeq \SI{0.8}{mV}.
\label{gammaI}
\end{align}
Using estimate (\ref{gammaI}) we find that $\chi_e\epsilon_0 \gamma^2 M_0^2 \simeq 3 \times 10^{-16} \mbox{\, J/m}$, which is much less than $\alpha M_0^2 \simeq 8\times 10^{-12} \mbox{\, J/m}$ whereas  is much less. This indeed justifies our simplification of the formula  (\ref{rho}) in Sec.\ref{Sec_Model}.
Now from (\ref{qe}) using $h=a\approx \SI{4}{\angstrom}$we find the vortex electric charge per one layer $q_{e} \approx 9 \times 10^{-22}\mbox{\,C} \approx 0.006 e$ ($e$ is the charge of electron). From (\ref{qm}) we find the value of ``magnetic charge'' and using (\ref{vcrit}) we find the critical tip voltage for creation of first VA pair $\varphi_0^{crit} \simeq 20 \mbox{\,V} \times \ln(H/a)$. Using (\ref{crit_charge}) with $\ln(H/a)\sim 1$ we also estimate the charge of impurity necessary to
create one VA pair in a monoatomic film layer as $Q^{crit}_{impurity} \simeq 10^{-18} \mbox{\,C} \approx 6e$.


\emph{Type-II multiferroics}. Namely, type-II multiferroics with a noncollinear magnetic order seem to be very natural candidates for observation of ``magnetic atoms''. Among them are cycloidal helimagnets such as $\mathrm{TbMnO_3}$, $\mathrm{DyMnO_3}$, $\mathrm{Eu_{0.75}Y_{0.25}MnO_3}$, $\mathrm{Ni_3 V_2 O_8}$, $\mathrm{Mn W O_4}$ etc., see \cite{Kimura:2007,Tokura:2014} for reviews.
Although the ground state in cycloidal helimagnets is different from the collinear one, vortex excitations of this ground state can be created anyway.

Here we make estimates for the prototypical type-II multiferroic $\mathrm{TbMnO_3}$. At room temperature it has orthorhombically distorted perovskite structure with space group $Pbnm$ and the lattice parameters $a = \SI{5.3}{\angstrom}$, $b = \SI{5.86}{\angstrom}$, $c = \SI{7.49}{\angstrom}$ [\onlinecite{Kenzelmann:2005}].
Below $T_N = 41\mbox{\,K}$ magnetic order appears -- sinusoidal spin density wave caused by frustrations is formed, with no electric polarization.
Below $T=28\mbox{\,K}$ magnetic cycloidal state in $bc$ easy plane with the wavevector $k=\pi(0,0.28,0)$ (and $\lambda = 2\pi/k \approx \SI{42}{\angstrom} \approx 7 b$) becomes energetically favorable and it induces the electric polarization which grows with lowering the temperature and reaches the value $P=8 \times 10^{-4}\,\SI{}{C/m^2}$
at $T=10\mbox{\,K}$; dielectric constants are $\epsilon_a \simeq 24$, $\epsilon_b \simeq 23$, $\epsilon_c \simeq 29$ [\onlinecite{Kimura:2003}], for estimates we get the mean value $\epsilon_{bc} = (\epsilon_b+\epsilon_c)/2 = 26 = 1+4\pi\chi_e$.
Using (\ref{PM}) we calculate the magnetoelectric constant
\begin{align}
\gamma M_0^2 = P/\chi_e \epsilon_0 k \simeq \SI{2.4}{mV}.
\label{gammaII}
\end{align}
As expected its value is greater than for type-I multiferroics (\ref{gammaI}), though the difference is not that drastic.
From (\ref{qe}) we also find the vortex electric charge per one layer $q_{e} \approx 1.8 \times 10^{-21}\mbox{\,C} \approx 0.01 e$.

In order to estimate the exchange stiffness and ``magnetic charge'' we use values of microscopic exchange constants.
Literature data on these is somewhat controversial, with even different relative strengthes of exchange constants along different unit-cell directions \cite{Xiang:2008,Nagaosa:2010,Milstein:2015,Fedorova:2015}.
For the estimate we use the exchange in the $c$-direction, which is free of magnetic frustrations, as $J_c=\SI{1.5}{meV}$, consistent with estimates from the Neel temperature \cite{Heffner:2001} $k_B T_N \simeq J z/3$ ($z$ is the number of nearest neighbors for the magnetic atom); therefore the exchange stiffness $A = J_c c^2/2V  \simeq \SI{0.3}{pJ/m}$.
Then from (\ref{vcrit}) we find the critical tip voltage for creation of first VA pair $\varphi_0^{crit} \simeq 0.3 \mbox{\,V} \times \ln(H/a)$. Using (\ref{crit_charge}) with $\ln(H/a)\sim 1$ we estimate the charge of impurity necessary to create one VA pair in a single layer as $Q^{crit}_{impurity} \simeq 1.6\times10^{-20} \mbox{\,C} \approx 0.1e$.


\emph{Other materials}.
In principle there might be a wide class of materials where magnetic vortices can be created by electric field, with multiferroics substituting only a narrow subclass of them: potentially, any magnetic insulator can exhibit a magnetoelectric phenomena, when symmetry arguments allow this \cite{Khomskii:2016}. One microscopic mechanism for this is the Dzyaloshinskii-Moriya interaction, which is present in many materials, such as weak ferromagnets, helical magnets, etc.


\subsection{Magnetic anisotropy}

Here we discuss the limitations of our theoretical picture due to possibility of out-of-plane polarization induced by the tip of the cantilever.
The charged tip induces an electric field perpendicular to the plane and consequently a transverse electric polarization, the latter can in turn induce a magnetic spiral with out-of-plane magnetisation component and in-plane wavevector.

In order this effect to be minor there should be some uniform easy-plane anisotropy that also holds magnetization in the plane in the absence of the tip. For example, single-ion anisotropy or one connected with the sample geometry (for ferromagnetic systems it is just the demagnetization energy; but for antiferromagnetic systems some weak surface anisotropy effects are also expected \cite{Caminale:2014}).

First, consider a system with tetragonal or orthorhombic symmetry, where an easy plane exists, and magnetic anisotropy energy is relatively big. Estimate the necessary transverse electric field that can induce a spiral in a ``hard'' plane (plane containing the hard direction). Here we chose as an example $\mathrm{TbMnO_3}$, which has a ground state with a cycloidal spiral in $bc$ plane (thus hard axis is $a$); for estimate we take anisotropy parameter $K_{an} \lesssim 0.1 J \simeq 0.1 \mbox{\, meV}$ [\onlinecite{Xiang:2008}] per one magnetic Mn atom.

Let the electric field be strong enough to induce a cycloidal magnetic spiral with an out-of-plane magnetization component. Using (\ref{w_m}) and (\ref{w_me}) we get the following estimate for the energy gained by the spiral state with induced out-of-plane polarization $P_z = \epsilon_0 \chi_e E_z = \epsilon_0 \chi_e E_{0z}/\epsilon \approx \epsilon_0 E_{0z}$ ($z$-axis is perpendicular to the plane, $E_{z}$ is the field inside film, $E_{0z}$ is the field in vacuum, $\epsilon \approx 1+\chi_e \approx \chi_e$):
\begin{align}
w_m + w_{me} \simeq -\gamma\epsilon_0 E_{0z} k M_0^2 + \frac{\alpha}{2} M_0^2 k^2
\label{w_mme1}
\end{align}
Minimizing this with respect to wavevector $k$ we obtain $k \simeq \gamma\epsilon_0 E_{0z}/\alpha$, therefore (\ref{w_mme1}) transforms to
\begin{align}
w_m + w_{me} \simeq - \frac{\gamma^2\epsilon_0^2 E_{0z}^2 M_0^2}{2\alpha}
\label{w_mme2}
\end{align}
which is the energy density that system gains when the spiral state is created. However the anisotropy energy density of the system increases by an amount:
\begin{align}
w_{anisotr} \simeq 4\times \frac{K_{an}}{2V},
\label{w_anisotr}
\end{align}
where factor of $4$ comes because of $4$ magnetic Mn atoms per $\mathrm{TbMnO_3}$ unit cell of volume $V=abc$ and factor of $2$ in the denominator because of averaging local anisotropy energies $\sim\cos^2 (kx)$ over spiral propagation direction $x$. Comparing the gained by the spiral state energy (\ref{w_mme2}) and the increase of the anisotropy energy (\ref{w_anisotr}), and taking typical values of parameters for $\mathrm{TbMnO_3}$  considered above, we obtain the following estimate for electric field, needed to produce a magnetic cycloid rotating in a ``hard'' plane
\begin{align*}
E_{0z} \gtrsim \frac{2\sqrt{\alpha K_{an}}}{\sqrt{V}\epsilon_0\gamma M_0} \simeq 1.9 \times 10^{10} \mbox{\,V/m}.
\end{align*}
For comparison, if the tip is placed at $H=\SI{10}{nm}$ above the film, in order to create such strong electric field that will influence $M_z$ component, the tip's voltage has to be as large as $\varphi = E_{0z} H \approx \SI{200}{V}$ (and even larger for greater $H$). This value is well above the critical voltage $\SI{0.3}{V}$ for VA pair formation estimated for $\mathrm{TbMnO_3}$ earlier.
Thus, we conclude that almost any reasonable nonzero magnetic easy-plane anisotropy protects the in-plane spin arrangement, in spite of the possibly strong perpendicular to the plane electric field component induced by the tip.

For systems with very weak or no single-ion anisotropy, such as $\mathrm{BiFeO_3}$, where only surface effects protect the in-plane magnetization, perpendicular component of the electric field will induce the out-of-plane component of the magnetization.

\begin{figure}[tbh]
\centering
  \includegraphics[width=0.6\linewidth]{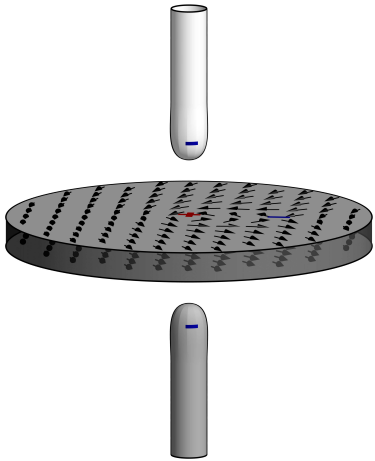}%
\caption{(Color online) Geometry of the system with two symmetric charged tips.}
\label{fig_geometry2}
\end{figure}

One way to avoid an out-of-plane magnetisation for such systems is to use two symmetrical charged tips from both sides of the film (Fig.\ref{fig_geometry2}; still $H \ll h$).

However we claim \cite{elsewhere} that even for single tip geometry (Fig.\ref{fig_geometry}), an out-of-plane component of magnetization near the tip would be localized inside the area of the radius $\sim H$ (distance from the tip to film surface).
This will not affect parts of the system outside the latter area, away from the tip. Therefore, the ``magnetic atom'' nucleus consisting of vortices in $r\sim H$ area may drastically change, but antivortex shells will preserve their qualitative features.
This will happen because of the topological nature of vortex excitations. For example, if anisotropy is not strong enough, vortices may have core with an out-of-plane component of ${\mathbf M}$ -- this will make the core electric charge more smeared, but it will not influence its amount. For a single vortex, the boundaries of the sample are negatively charged with the same charge as if the vortex core was ``flat''. The boundaries are not influenced by an out-of-plane component of the electric field of the tip, since it decreases as $\sim 1/r^3$ away from the center, while in-plane field decreases only as $\sim 1/r^2$. Then by the charge conservation, the positive charge at the vortex core has to remain the same.


\section{Conclusions}

In magnetoelectric materials interaction of magnetic and electric subsystems makes possible the creation of magnetic patterns by an applied electric field. In this paper we have presented analytical and numerical analysis of such quasi-two-dimensional type-II multiferroic-like materials, where ``atom''-like patterns of magnetic vortices and antivortices can be created by a strong electric field,
and studied structure and electric polarizability of such magnetic ``atoms''.

When the spot-like electric field, produced by a tip of a cantilever, exceeds some critical value, the first vortex-antivortex pair forms; with increasing electric field more vortices and antivortices are created from the ``magnetic vacuum'' of the magnetoelectric film. We found analytically and confirmed numerically the critical voltage that is required to create the first vortex-antivortex pair, and found the dependence of the number of vortex pairs on the applied voltage. We have shown that a ``magnetic atom'' forms: a ``nucleus'' consists of vortices arranged in a lattice-like structure with a short range order; it is surrounded by antivortices forming outer concentric circular ``shells''. We obtained the vortex density profile in the continuous approximation.

In the studied here type-II multiferroic-like magnetoelectric materials vortices carry a positive and antivortices -- a negative electric charge, therefore additional weak in-plane electric field pulls them in the opposite directions, thus creating a polarization. In this work we have studied the temperature dependence of electric polarizability of such ``magnetic atoms''.
Polarizability of finite vortex-antivortex cluster is determined by the eigenvalues of its Hessian matrix.
We analyzed properties of these normal modes using results from Euclidean random matrix theory. We speculate that the behavior of polarizability $\alpha \sim 1/T^{1-\eta}$ is generic for disordered (``glassy'') finite classical systems with many local minima near the ground state energy; the exponent $\eta$ depends on the form of the potential and dimensionality of the system. Thermodynamical fluctuations wash out the fine details of the energy landscape, effectively cutting off the low-lying eigenmodes. By the exact numerical diagonalization of Hessian matrix we get the density of its eigenvalues, from which we deduce $\alpha \sim 1/T^{1-\eta}$ ($\eta \approx 0.25$) dependence of polarizability and confirm it directly by Monte Carlo simulation.

The presented theoretical and numerical analysis should find its experimental realization.
First, such ``magnetic atoms'' can be created in thin films of type-II multiferroic materials with easy-plane spin arrangement, for example in cycloidal helimagnets (such as $\mathrm{TbMnO_3}$, $\mathrm{Ni_3 V_2 O_8}$  and others\cite{Tokura:2014}). Although the ground state in cycloidal helimagnets is different from the collinear one, vortex excitations upon this ground state can be created anyway. Also, the materials where magnetic vortices can be created by electric field, are not restricted to only multiferroics: potentially, any magnetic insulator can exhibit a magetoelectric phenomena, when symmetry arguments allow this.

Atomic force microscopy (AFM) techniques might be used in order to create high local electric field.
The direct observation of the magnetic vortex patterns formation can be done using high-resolution Lorentz TEM imaging \cite{Yu:2010,Ratcliff:2016} combined with AFM \cite{Erts:2001}; alternatively combined AFM/STM setups can be used \cite{Nakayama:2012,Hapala:2015}.
A number of indirect methods can be used in order to test the eigenmodes of the induced vortex-antivortex system. These eigenmodes can be experimentally found by measuring the electric polarization in the applied transverse electric field (as studied analytically in Sec.\ref{Sec_polarizability_RMT} and numerically in Sec.\ref{Sec_Numerical}), but also by measuring heat capacity, or microwave absorption spectra experiments \cite{Onose:2012}.
Dynamical effects associated with the vortex ``atom'' can be tested using the setup, when the cantilever oscillates along the film plane \cite{Yazdanian:2009} or applying a time-dependent voltage, and measuring the energy dissipation.

We finally note that the studied here electric field creation of magnetic vortices and antivortices out of magnetoelectric thin film ground state provides a condensed-matter toy-model for studying the long-standing problem of production of electron-positron pairs out of vacuum applying strong electric fields, which arises in quantum electrodynamics and atomic physics.


\begin{acknowledgements}
The authors are grateful to S.A. Brazovskii for the idea to consider field-induced topological superlattices; to D.I. Khomskii and M.V. Mostovoy for helpful discussions.

We acknowledge the financial support of the Ministry of Education and Science of the Russian Federation in the framework of Increase Competitiveness Program of NUST MISiS (No K2-2016-067).
\end{acknowledgements}

\appendix

\section{Energy of a vortex in the in-plane electric field}

\begin{figure}
\subfloat[Distribution of polarization for single vortex. \label{disk}]{%
  \includegraphics[width=.45\linewidth]{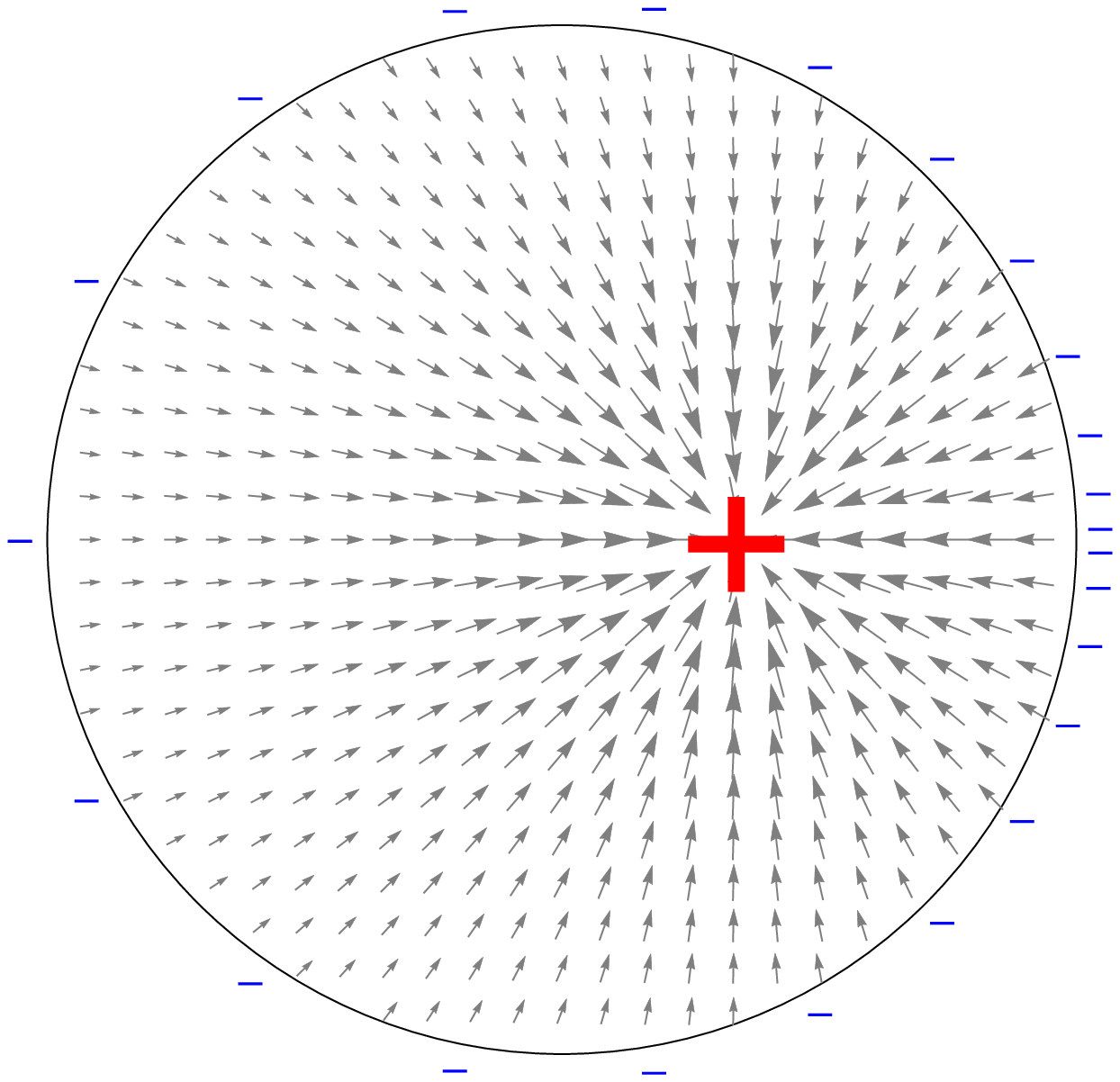}%
}
\hfill
\subfloat[Coordinate system. \label{coord}]{%
  \includegraphics[,width=.51\linewidth]{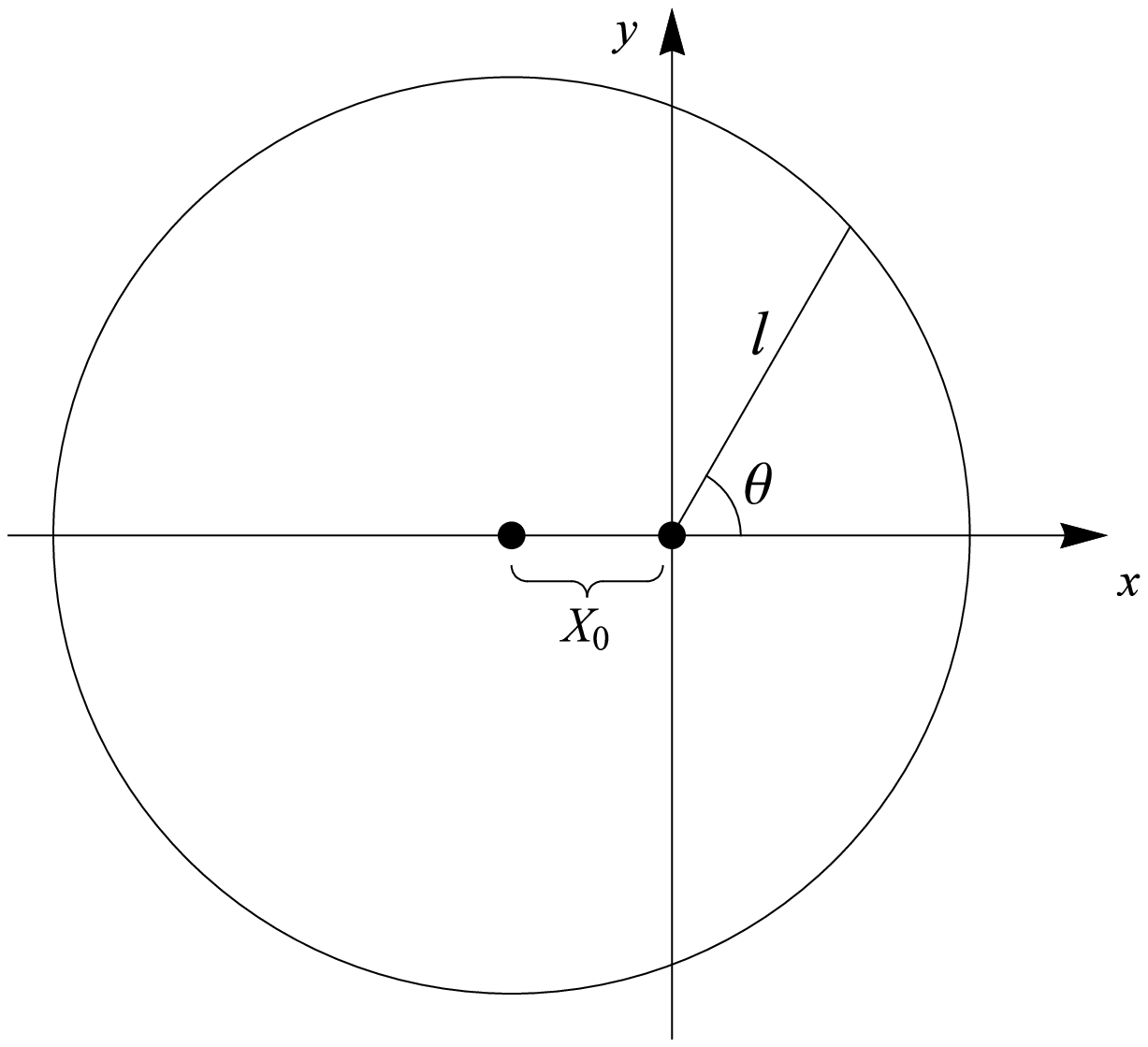}%
}
\caption{(a) Distribution of polarization for single vortex. Vortex core is positively charged, edge of the sample is negatively charged.
         (b) Coordinate system. Vortex core is placed at $(0,0)$; disk center is $(-X_0,0)$. }
\label{figappendix}
\end{figure}

In this Appendix we calculate the energy of a charged vortex in the electric field. As it was explained in the main text, the magnetic vortex core acquires an electric charge (\ref{qe}) due to magnetoelectric coupling (\ref{energy-density}). Since our sample is electrically neutral, its edge acquires negative charge of the same magnitude, which should also be taken into account (see Fig.\ref{disk}).
This edge charge effectively shields the vortex charge, modifying its energy.

Let our sample be disk with radius $R$, which contains a single vortex with vorticity $n$, placed at the distance $X_0$ from the center of the disk. Chose a coordinate system as shown in Fig.\ref{coord}; electric field ${\mathbf E} = (E_x, E_y)$.

From (\ref{polarization}) polarization of single vortex configuration (\ref{vortex}) is
\begin{equation}
\label{P-app}
{\mathbf P} = \gamma\chi_e M_0^2  \left(
\begin{array}{ccc}
- \partial_y\phi \\
  \partial_x\phi \\
 \end{array}
\right) =
-\frac{k\gamma\chi_e M_0^2}{r}  \left(
\begin{array}{ccc}
  \cos\theta \\
  \sin\theta \\
 \end{array}
\right).
\end{equation}
Since we calculate the linear response, $\chi_e {\mathbf E}$ term was omitted from (\ref{polarization}).
Then the electric energy of magnetic vortex configuration is $W_{me} = - h \int {\mathbf P} {\mathbf E} \, d^2 r$
\begin{equation}
W_{me} = 
 k h\gamma\chi_e M_0^2 \int_0^{2\pi} d\theta  (E_x \cos\theta + E_y \sin\theta) l(\theta).
\end{equation}
Since $l(\theta) = \sqrt{R^2 - X_0^2 \sin^2\theta} -X_0 \cos\theta \,$ is an even function of $\theta$, then the odd term $E_y\sin\theta$ vanishes. Thus,
\begin{align}
W_{me} =  
        -\pi k h\gamma\chi_e M_0^2 E_x X_0.
\end{align}

Using (\ref{qe}) we obtain $W_{me} = -\frac{1}{2} q_e  E_x X_0$ or for general displacement of vortex core ${\mathbf r} = (X_0, Y_0)$
\begin{equation}
\label{energy-app}
W_{me} = -\frac{1}{2} q_e  {\mathbf E} {\mathbf r},
\end{equation}

We see that for the circular sample geometry this energy is exactly a half of the vortex core energy of charged particle in the in-plane electric field. This effective shielding comes from the boundary charge and its exact value depends on the geometry of the sample.
We also note that since polarization (\ref{P-app}) is linear in phase angle $\phi$, the same result (\ref{energy-app}) holds for vortex-antivortex configuration, with ${\mathbf r}$ denoting a vector that connects cores of the antivortex and the vortex.

\section{Thermally activated pairs}

In this Appendix we find the polarizability that comes from thermally activated dipole pairs at $T < T_{BKT}$. Since the chemical potential of a pair is quite big, the concentration of pairs is low and we can employ the approximation of non-interacting pairs. Therefore we can find polarizability of one pair, then multiply it by the number of pairs and we will get polarizability  of the VA pairs subsystem.

First, find the polarizability of one VA dipole pair.
Energy of a dipole pair is the sum of magnetic interaction energy of the vortex and antivortex (\ref{Wm}) and the magnetoelectric energy (\ref{energy-app}), obtained in the Appendix A.
\begin{equation}
\label{E1pair}
W  = 2q_m^2 \ln(r/a) - q_e E x/2
\end{equation}
Here ${\mathbf r} = (x, y)$ is the vector connecting the antivortex core to the vortex core; the electric field is chosen to be parallel to the $x$-axis.
The partition function of such a dipole:
\begin{equation*}
Z(E) =  \frac{S}{a^2} \int\frac{d^2 r}{a^2} \exp\left( -\beta (2q_m^2 \ln(r/a) + q_e E x/2) \right)
\end{equation*}
where $S=\pi R^2$ is the area of the system, which comes from integration over the dipole center of mass. For zero electric field we get
\begin{align*}
Z(0) \equiv Z_0 = \frac{S}{a^2}\frac{\pi}{\beta q_m^2-1}
\end{align*}

Average dipole moment of a pair:
\begin{align*}
p(E) = \frac{S}{Z(E) a^2} \int\frac{d^2 r}{a^2} \frac{q_e}{2} x \exp\left( \beta (-2q_m^2 \ln(r/a) + q_e E x/2) \right)
\end{align*}
Formally, this integral diverges at large $r$. However, we should keep in mind that we are calculating
linear response (i.e., we take the limit $E \rightarrow 0$), so that $r$-divergence is cut off at the radius of the sample
$r = R$. Therefore, if we first take derivative of $p(E)$ with respect to $E$ and then put $E = 0$,
then we obtain the correct result.
Hence, polarizability of one pair is
\begin{align*}
\alpha_1 = \left.\frac{\partial p}{\partial E} \right|_{E=0} =
\frac{\beta q_e^2 S}{4 Z_0 a^2} \int\frac{d^2 r}{a^2} x^2 \exp\left( -2\beta q_m^2 \ln(r/a) \right),
\end{align*}
integrating which we obtain
\begin{align}
\alpha_1 = \frac{\beta q_e^2 a^2}{8} \frac{\beta q_m^2-1}{\beta q_m^2-2}
\label{alpha_1}
\end{align}

Second, estimate the concentration of VA pairs. Since a weak field does not affect their concentration, we make the estimation for $E=0$. Moreover, due to the fact that the chemical potential of a pair is quite big with respect to $T_{BKT}$: ($\mu \simeq -\pi q_m^2$, $T_{BKT} \simeq q_m^2/2$,
see Ref.[\onlinecite{KT1}]), concentration of VA pairs is always small. Calculate the grand partition function of non-interacting dipole system:
\begin{equation*}
\label{grand0}
\begin{split}
\mathcal{Z} 
&=  \sum_{n} \exp(\beta\mu n) \sum_{{{\text{different} \atop \text{config}}}}
\exp\left( -\beta(W({\mathbf r_1})+... + W({\mathbf r_n})) \right) =  \\
&=\sum_{n} \frac{1}{n!}\left(\exp(\beta\mu) \frac{S}{a^2} \int\frac{d^2 r}{a^2} \exp\left( -\beta W({\mathbf r}) \right)\right)^n = \\
&= \exp(e^{\beta\mu} Z_0).
\end{split}
\end{equation*}

Then, the number of thermally activated VA pairs
\begin{align}
N_{pairs} = \frac{T}{\mathcal{Z}} \frac{\partial \mathcal{Z}}{\partial \mu} = e^{\beta\mu} Z = \frac{\pi S e^{\beta\mu}}{(\beta q_m^2-1) a^2}
\label{Npairs}
\end{align}
Finally, combining (\ref{alpha_1}) and (\ref{Npairs}) we find the polarizability of the sample that comes from all dipole pairs
\begin{align}
\alpha_{thermal} = N_{pairs} \alpha_1 = \frac{\pi^2 R^2 \beta q_e^2 e^{\beta\mu}}{4(\beta q_m^2-2)}
\label{alpha_quasi_exact}
\end{align}
From here we can deduce the electric susceptibility $\chi_e = \alpha_{thermal}/S$ and we see that it is consistent with Kosterlitz-Thouless's susceptibility \cite{KT1} with the difference that now we have the electric charge $q_e/2$ instead of $q_m$ in the numerator. This is natural since the response to
the electric field depends on the value of electric charge $q_e$ whereas screening comes from the interaction of "magnetic" charges $q_m$.

We see that at low temperatures $T\rightarrow 0$, $\alpha_{thermal}$ exponentially vanishes to $0$, whereas diverges as $\alpha_{induced} \sim 1/T^{1-\eta}$, therefore $\alpha_{induced}$ dominates. On the contrary, near $T_{BKT}$ $\alpha_{thermal}$ diverges and dominates over $\alpha_{induced}$.
\\
\\


\begin{thebibliography}{25}
\bibitem{elsewhere} the details will be presented elsewhere
\end{thebibliography}
\end{document}